\documentclass[a4paper,11pt]{article}

\pdfoutput=1 % if you are submitting a pdflatex (i.e. if you have
\usepackage{jheppub,bm,slashed} 
\usepackage[legacycolonsymbols]{mathtools}
\usepackage{xcolor}
\usepackage{extarrows}
\usepackage{easybmat}

\newcommand{\blue}[1]{\textcolor{blue}{#1}}

\usepackage{nicematrix}
\usepackage[T1]{fontenc} % if needed
\usepackage[utf8]{inputenc}

\renewcommand*{\backref}[1]{}
\renewcommand*{\backrefalt}[4]{%
  \ifcase #1 %
    \relax % use \relax if you do not want the "No citations" message
  \or
    {\scriptsize (page~#2).}%
  \else
    {\scriptsize (pages~#2).}%
  \fi%
}

\definecolor{light_blue}{rgb}{0.15, 0.35, 0.95}
\definecolor{kit_green}{rgb}{0, 
0.58823 %150/255
, 0.50980 %130/255
}

\newcommand{\tchar}[2]{%
  \vartheta\genfrac{[}{]}{0pt}{}{#1}{#2}%
}

\newcommand{\tbchar}[2]{%
  \overline{\vartheta}\genfrac{[}{]}{0pt}{}{#1}{#2}%
}

\newcommand{\chvec}[2]{%
  \genfrac{[}{]}{0pt}{}{#1}{#2}%
}

\usepackage{array}
\usepackage{amsmath}
\usepackage{arydshln}
\usepackage{tikz}
\usepackage{pgfplots}
\pgfplotsset{compat=1.14}
\usepackage{tikz-3dplot}
\usetikzlibrary{intersections, calc,positioning,decorations.pathreplacing,decorations.pathmorphing,arrows,arrows.meta,patterns,pgfplots.fillbetween,math,matrix}

\usepackage{newtxtext}
\usepackage{bm}
\usepackage{amsfonts}
\usepackage{amssymb}
\usepackage{latexsym}
\usepackage{graphicx}
\usepackage{float}
\usepackage{siunitx}
\usepackage{booktabs}
\usepackage{braket}
\usepackage{physics}
\usepackage{cancel}
\usepackage{ascmac}
\usepackage{comment}

\newcommand{\C}{\mathbb{C}}

\renewcommand{\H}{\mathcal{H}}

\newcommand{\N}{\mathcal{N}}
\newcommand{\R}{\mathbb{R}}

\newcommand{\M}{\mathcal{M}}
\newcommand{\Z}{\mathbb{Z}}

\renewcommand{\Im}{\mathrm{Im}}

\newcommand{\Spin}{\mathrm{Spin}}
\newcommand{\SO}{\mathrm{SO}}

\newcommand{\SU}{\mathrm{SU}}
\newcommand{\U}{\mathrm{U}}

\newcommand{\q}{q^{L_0}\bar q^{\bar{L}_0}}

\newcommand{\1}{\mathbf{1}}

\usepackage[whole]{bxcjkjatype} 
\usepackage{multirow}
\usepackage{booktabs} %for \toprule
\usepackage{tabularx} %for tabularx
\usepackage{tikz-cd}

\preprint{
\begin{minipage}{5cm}
\flushright
KEK-TH-2797
\end{minipage}}

\title{Non-Supersymmetric String-String Dualities via Enriques Surfaces}

\author[1]{Arata Ishige}

\affiliation[1]{Graduate Institute for Advanced Studies, SOKENDAI, 1-1 Oho, Tsukuba, Ibaraki 305-0801, Japan}

\emailAdd{arata@post.kek.jp}

\abstract{We propose non-supersymmetric analogues of $6d~\N=2$ Type~II/heterotic dualities via a quotient of a K3 surface: an \textit{Enriques surface}. We start from Type~II strings on a K3 surface and construct orbifold theories using an involution of K3. We extract the massless and tachyonic spectra and identify the moduli spaces locally. We further reinterpret the constructions as Type 0A/0B strings compactified on an Enriques surface, and argue that the theories are dual to recently constructed non-supersymmetric heterotic asymmetric orbifolds.}

\begin{document}
\maketitle

\section{Introduction}
Over the past several decades, the five superstring theories with spacetime supersymmetry have each played distinct and complementary roles, including the emergence of exceptional gauge symmetries, extended objects such as branes, and the advent of M-theory \cite{Witten_1995,GREEN1984117,Gross:1985fr,Gross:1985rr,Horava_1996_1,Horava_1996_2,CANDELAS198546,Maldacena:1997re,witten1995commentsstringdynamics,Aspinwall:1994rg}. 

These developments are unified by various dualities \cite{Vafa_1996,Morrison:1996na,Morrison:1996pp,Polchinski:1995mt,Polchinski_1996,aspinwall1999k3surfacesstringduality,Schwarz:1995bj,Bershadsky:1998vn}. In supersymmetric dualities, K3 surfaces play a crucial role by providing a bridge between Type II strings and heterotic strings. It has long been believed that the heterotic string on $T^4$ is dual to Type IIA superstring theory compactified on a K3 surface. It is also known that the strong-coupling limit of the heterotic string on $T^5$ is described by Type IIB theory compactified on a K3 surface. These two pairs share the same moduli space, massless spectrum, and objects.

While supersymmetric superstring theory has produced a series of remarkable results, non-supersymmetric string theories have also made steady, incremental advances \cite{Dixon:1986iz,SEIBERG1986272,Kawai:1986vd,Bergman:1999km,ALVAREZGAUME1986155,Klebanov_1999,Sugimoto_1999,sagnotti1995propertiesopenstring,Sagnotti_1997,Fabinger_2000,hellerman2008stablevacuumtachyonice8,Horava_2008}. Beyond the phenomenological motivation provided by the absence of observed supersymmetry in nature, non-supersymmetric settings have recently drawn renewed interest \cite{Kaidi_2021,nakajima2023newnonsupersymmetricheteroticstring,fraiman2025symmetriesdualitiesnonsupersymmetricchl,Hamada:2024cdd,hamada20258dnonsupersymmetricbranesheterotic,DeFreitas:2024ztt,Boyle_Smith_2024,Rayhaun_2024,hohn2024classificationselfdualvertexoperator,Saxena:2024eil,Koga:2022qch,Itoyama:2019yst,Itoyama:2021itj,Baykara:2024tjr}, in particular from the perspective of quantum gravity \cite{Ooguri:2016pdq,mcnamara2019cobordismclassesswampland,GarciaEtxebarria:2020xsr,Basile:2023knk,debray2023chroniclesiibordiadualitiesbordisms,Dierigl_2023,heckman2026gsodefectsiiaiibwalls,Kaidi:2023tqo,Kaidi:2024cbx}. Without the tight constraints of supersymmetry, the landscape of theories proliferates, even as the web of interrelations among them grows increasingly sparse. More theories, fewer relations. Can we find a framework that allows us to survey with clarity, in particular, non-perturbative dualities? Some examples were already proposed in \cite{Bergman:1997rf,Bergman:1999km,Blum:1997cs,Blum:1997gw,Blumenhagen:1999ad}, and more recently in \cite{fraiman2025symmetriesdualitiesnonsupersymmetricchl}. Aside from dualities, a systematic approach to classify non-supersymmetric heterotic strings in terms of chiral CFTs was begun in \cite{DeFreitas:2024ztt}.

\begin{figure}[h]
    \centering
    \begin{tikzcd}[row sep=huge, column sep=huge]
\text{Type IIA on }K3 
  \arrow[rr, leftrightarrow, "\mathrm{dual}"] 
  \arrow[d, "h"']
&&
\text{Het on }T^4 
  \arrow[d, "h"]
\\
\text{Type 0A on Enriques} 
  \arrow[rr, leftrightarrow, "\mathrm{dual}"] 
&&
\text{Het on }T^4/\mathbb{Z}_2
\end{tikzcd}

    \begin{tikzcd}[row sep=huge, column sep=huge]
\text{Type IIB on }K3 
  \arrow[rr, leftarrow, "\mathrm{strong~ coupling ~limit}"] 
  \arrow[d, "h"']
&&
\text{Het on }T^5
  \arrow[d, "h"]
\\
\text{Type 0B on Enriques} 
  \arrow[rr, leftarrow, "\mathrm{strong~ coupling ~limit}"] 
&&
\text{Het on }T^5/\mathbb{Z}_2
\end{tikzcd}
    \caption{A schematic picture. }
    \label{fig:scheamtic}
\end{figure}
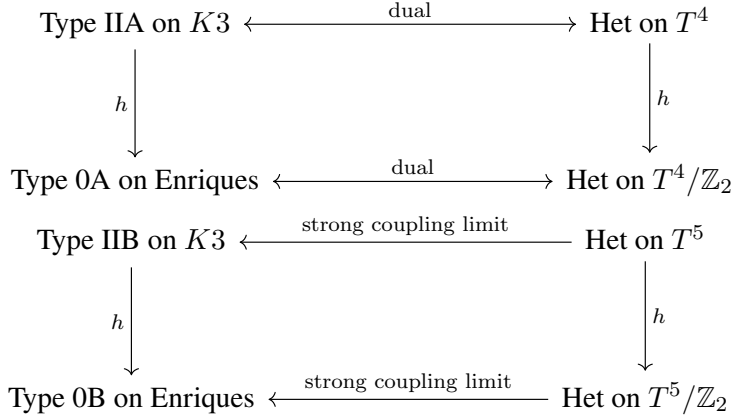

Another possible answer is offered by K3 surfaces and their quotients, known as \textit{Enriques surfaces}. We start from the string-string dualities between Type II strings and heterotic strings via K3 surfaces \cite{aspinwall1999k3surfacesstringduality,Witten_1995,Kiritsis_2000}. Both theories share the same moduli space, massless spectrum, and gauge symmetries. Further more, there is a common discrete symmetry $h$. It can be viewed as a $\Z_2$ automorphism of the lattice $\Gamma_{3,19}$ in the heterotic string \cite{Acharya:2022shu}, while it is encoded in the geometry of $K3$ surfaces as a $\Z_2$ involution on the Type II side.  Then, taking the quotient of both theories leads to a pair of new non-supersymmetric theories that nevertheless remain related by duality.  In this paper, we propose non-supersymmetric analogues of the string-string dualities (Figure \ref{fig:scheamtic}).

This paper is organized as follows. In Section~\ref{sec:het}, we review the supersymmetric heterotic strings and the non-supersymmetric asymmetric orbifold proposed in \cite{Acharya:2022shu}. In Section~\ref{sec:K3}, we review Type II compactification on a $K3$ surface and the corresponding supergravity description. In Section~\ref{sec:Enr}, we construct the non-supersymmetric orbifolds of Type II strings relevant to Enriques surfaces, and identify their moduli spaces and massless spectra. In Section~\ref{sec:Type 0}, we present an alternative viewpoint in terms of Type 0 strings, which would be useful to investigate subtle features of the theories. Finally, in Section~\ref{sec:dualities}, we discuss the non-supersymmetric analogues of Type II/heterotic dualities.

\section{Heterotic strings}\label{sec:het}

In this section, we review supersymmetric heterotic strings \cite{Gross:1985fr,Gross:1985rr,NARAIN198641} and a non-supersymmetric heterotic string constructed in \cite{Acharya:2022shu}.

\subsection{Heterotic strings on \texorpdfstring{$T^d$}{Td} (supersymmetric)}

The Hilbert space of heterotic strings on $T^d$ is given by
\begin{equation}
\H=\H_{\text{Boson}}^{8-d,8-d}\otimes \H_{\text{Fermion}}^{8,0}\otimes\H_{\Gamma_{d,d+16}},
\end{equation}
with the GSO projection. Here $\H_{\text{Boson}}^{8-d,8-d}, \H_{\text{Fermion}}^{8,0}$  are the Fock spaces for bosonic/fermionic oscillators. Remaining $\H_{\Gamma_{d,d+16}}$ is the Hilbert space associated with a lattice $\Gamma_{d,d+16}=\Gamma_{0,16}\oplus H_1(T^d;\Z)\oplus H^1(T^d,\Z)$, which is given by

\begin{equation}
    \H_{\Gamma_{d,d+16}}=\H_{\text{Boson}}^{d,d+16}\otimes \qty(\bigoplus_{p\in\Gamma_{d,d+16}^\ast}\C\ket{p}),
\end{equation}
where $\Gamma_{d,d+16}^\ast$ is the dual lattice of $\Gamma_{d,d+16}$. Its torus partition function is 

\begin{equation}
   Z= \frac{1}{(\Im \tau)^{\frac{8-d}{2}} \eta^{12} \bar\eta^{24}}
\left( \sum_{p\in V} - \sum_{p\in \mathrm{Sp}} \right)
q^{\tfrac12 p^2}
\sum_{P\in\Gamma_{d,d+16}^\ast}
q^{\tfrac12 P_L^2} \bar q^{\tfrac12 P_R^2},
\\
\end{equation}
where $V,Sp$ are vector or spinor conjugacy classes of $\SO(8)$ respectively (see Appendix \ref{subsec:characters}). Modular invariance imposes the condition that the lattice $\Gamma_{d,d+16}^\ast$ must be even and self-dual. The classification of $(d,d+16)$ even self-dual lattice is well-known: for $d=0$, there are two distinct choices: the $E_8\times E_8$ root lattice or the $\Spin(32)/\Z_2$ lattice. For $d\geq 1$, it is unique up to $\SO(d,d+16)$.

The moduli space of heterotic strings on $T^d$ is locally given by \cite{NARAIN198641}
\begin{equation}
\mathcal{M}^{\text{Het}}_{T^d}=\frac{O(d,d+16)}{O(d)\times O(16+d)}\times \R_{\text{dilaton}}.
\end{equation}
In other words, there are $\dim \mathcal{M}^{\text{Het}}_{T^d}=d(d+16)+1$ massless scalar fields in the theory. $d^2+1$ components come from the metrics and B-fields on $T^d$, and $16d$ components come from the Wilson lines (rank $16$) along $T^d$.

\subsection{An automorphism of the lattice}
We focus on a $\Z_4$ symmetry $h$ of the heterotic string \cite{Acharya:2022shu}. First, we choose the following representation of the lattice:

\begin{equation}
\begin{aligned}
        \Gamma_{d,d+16}=&E_8\oplus E_8\oplus \Gamma_{d,d}\\
        =&E_8\oplus E_8\oplus \Gamma_{1,1}\oplus\Gamma_{1,1}\oplus\Gamma_{1,1}\oplus\Gamma_{d-3,d-3},
\end{aligned}
\end{equation}
here we assume $d\geq 3$. In this representation we set the Wilson lines and the B-fields to zero and take the metric of $T^d$ to be diagonal.

Next, we consider the following $\Z_2$ action of $h$ on the lattice:

\begin{equation}
\begin{aligned}
        h:&\qty(x_1^{(E_8)},x_2^{(E_8)},y_1,y_2,y_3,y_{d-3,d-3})\\
        \mapsto&\qty(x_2^{(E_8)},x_1^{(E_8)},y_2,y_1,-y_3,y_{d-3,d-3}),
\end{aligned}
\end{equation}
The decomposition of the lattice under $h$ is given by
\begin{equation}
    \Gamma_{d,d+16}=I_{d-2,d+6}\oplus N_{2,10},
\end{equation}
where
\begin{equation}
\begin{aligned}
        I_{d-2,d+6}=&\{x\in\Gamma_{d,d+16}| h(x)=x\}\\
        =&\sqrt{2}E_8\oplus \sqrt{2}\Gamma_{1,1}\oplus\Gamma_{d-3,d-3},\\
        N_{2,10}=&\{x\in\Gamma_{d,d+16}| h(x)=-x\}\\
        =&\sqrt{2}E_8\oplus \sqrt{2}\Gamma_{1,1}\oplus\Gamma_{1,1}.\\
\end{aligned}
\end{equation}
The moduli space of this decomposition is given by
\begin{equation}
    \frac{O(d-2,d+6)}{O(d-2)\times O(d+6)}\times\frac{O(2,10)}{O(2)\times O(10)}\qty(\subset \frac{O(d,d+16)}{O(d)\times O(16+d)}),\quad d\geq 3
\end{equation}
and the dimensions are $(d-2)\cdot(d+6)+2\cdot10 =(d+2)^2+4$. In \cite{Acharya:2022shu}, $\frac{O(d-2,d+6)}{O(d-2)\times O(d+6)}$ is called Coulomb branch, while $\frac{O(2,10)}{O(2)\times O(10)}$ is called Higgs branch.

\subsection{Asymmetric orbifold (non-supersymmetric)}
We now consider an asymmetric orbifold of heterotic strings by the lattice automorphism $h$, which corresponds to the case of $(s,a,\delta)=(10,10,0)$ in \cite{Acharya:2022shu}. 
First, we must specify the action of $h$ on the full Hilbert space.

Up to exchanging two $E_8$'s, $h$ can be viewed as an element of $\SO(4)\times \SO(4)\subset \SO(8)$ on $\R^4\times T^4$ as follows:

\begin{equation}
    h=\qty(\begin{matrix}
        1_{4}&&&\\
        & &1 &\\
        &1&&\\
        &&&-1\\
        &&&&1
    \end{matrix})\in \SO(8).
\end{equation}
Since this is a $\pi$ rotation, the relation $h^2=1\in \SO(8)$ lifts to $h^2=-1\in\Spin(8)$. Then $h^2$ acts as $(-1)^F$ on the Hilbert space, where $F$ is the spacetime fermion number. Therefore, while $h$ is a $\Z_2$ involution on the lattice, it is a $\Z_4$ symmetry on the Hilbert space. To make the partition function modular invariant, $h$ should also act on $\H_{\Gamma_{d,d+16}}$.

The partition function and its building blocks are given as follows:

\begin{equation}
\begin{aligned}
        Z=&\frac{1}{4}\sum_{a,b=0}^3 Z_{a,b},\\
        Z_{a,b}=&\tr_{\H^{h^a}}h^b \q,
\end{aligned}    
\end{equation}
where $\H^{h^a}$ is the $h^a$-twisted Hilbert space. Actually one does not need to construct twisted Hilbert spaces, because these building blocks $Z_{a,b}$ are connected to each other by the modular transformations as follows:

\begin{equation}
    \begin{aligned}
        T\cdot Z_{a,b}=Z_{a,a+b},\\
        S\cdot Z_{a,b}=Z_{b,-a}.\\
    \end{aligned}
\end{equation}
As a starting point, partition functions in the untwisted sector are given as follows:

\begin{equation}
\begin{aligned}
       Z_{0,0} = &\frac{1}{(\Im \tau)^{\frac{8-d}{2}} \eta^{12} \bar\eta^{24}}
\left( \sum_{p\in V} - \sum_{p\in \mathrm{Sp}} \right)
 q^{\tfrac12 p^2}
\sum_{P\in\Gamma_{d,d+16}}
q^{\tfrac12 P_L^2} \bar q^{\tfrac12 P_R^2},
\\
Z_{0,2n+1} =&\frac{1}{(\Im\tau)^{2}  \eta^{12} \bar\eta^{24}}
\left( \frac{2\eta^3}{\vartheta_{10}} \right)^{5}
\left( \frac{2\bar\eta^3}{\bar\vartheta_{10}} \right)
\left( \sum_{p\in V} - \sum_{p\in \mathrm{Sp}} \right)
q^{\tfrac12 p^2} e^{-2\pi i p\cdot(2n+1) v_f}\\
&\times\sum_{P\in I_{6+d,d-2}}
q^{\tfrac12 P_L^2} \bar q^{\tfrac12 P_R^2}
e^{2\pi i P\cdot(2n+1) v},\\
    Z_{0,2} =& \frac{1}{(\Im\tau)^{2} \eta^{12} \bar\eta^{24}}
\qty( \sum_{p\in V} - \sum_{p\in \mathrm{Sp}} )
 q^{\tfrac12 p^2} e^{-4\pi i p\cdot v_f}
\sum_{P\in\Gamma}
e^{4\pi i P\cdot v} e^{2\pi i P_I^2}
q^{\tfrac12 P_L^2} \bar q^{\tfrac12 P_R^2}.
\end{aligned}
\end{equation}
Here,  $v_f=\qty(\frac{1}{2},0,0,0)$ represents the action of $h$ on $\H_{\text{Fermion}}^{8,0}$, and one can check that $h^2$ plays the same role as $(-1)^F$:

\begin{equation}
    \qty(e^{2\pi ip\cdot v_f})^2=\begin{cases}
        1,\text{for } p\in V,\\
        -1, \text{for } p \in Sp.
    \end{cases}
\end{equation}
From these functions, other partition functions can be obtained as follows:
\begin{equation}
    \begin{aligned}
        Z_{a,0}=&S\cdot Z_{0,a},\\
        Z_{1,n}=&T^n \cdot Z_{1,0},\\
        Z_{2,2n-1}=&T^n\cdot S\cdot  Z_{1,2},\\
        Z_{2,2}=&T\cdot Z_{2,0},\\
        Z_{3,3n-1}=&T^n\cdot S\cdot Z_{1,3}.
    \end{aligned}
\end{equation}
For example,

\begin{equation}
\begin{aligned}
Z_{1,n} = &\frac{1}{(\Im\tau)^{2}\eta^{24} \bar\eta^{12}}
\frac{1}{\sqrt{|I^*/I|}}
\qty( \frac{2\eta^3}{\vartheta\chvec{0}{n+1}} )^{5}
\qty( \frac{2\bar\eta^3}{\bar\vartheta\chvec{0}{n+1}} )
\\
\times
\left( \sum_{p\in V} - \sum_{p\in \mathrm{Sp}} \right)&
\bar q^{\tfrac12 (p+v_f)^2}e^{-\pi i n (p+v_f)^2}
\sum_{P\in I^\ast_{d-2,d+6}}
e^{\pi i n (P+v)^2}
q^{\tfrac12 (P+v)_L^2}
\bar q^{\tfrac12 (P+v)_R^2}.
\end{aligned}
\end{equation}

\subsection{Spectrum}

Following \cite{Acharya:2022shu}, the spectrum of the asymmetric orbifold can be organized by first decomposing the Narain lattice into the $h$-invariant and $h$-anti-invariant sublattices,
\begin{equation}
    \Gamma_{d,d+16}=I_{d-2,d+6}\oplus N_{2,10},
\end{equation}
which in turn motivates the local factorization of the moduli space into the so-called Coulomb and Higgs branches:
\begin{equation}\label{eq:het_moduli}
   \M^{\text{Het}}_{T^d/h}= \frac{O(d-2,d+6)}{O(d-2)\times O(d+6)}\times\frac{O(2,10)}{O(2)\times O(10)}\times \R_{\text{dilaton}}, \quad d\geq3,
\end{equation}
which is $(d+2)^2+5$-dimensional. 

In particular, one expects:
(i) a generic point in moduli space has an abelian gauge group (the Cartan subalgebra), while
(ii) at special loci additional lattice vectors become massless and enhance the gauge symmetry.

%At the level of the worldsheet construction, the non-supersymmetric orbifold is consistent because the building blocks $Z_{a,b}$ obey the modular transformation rules in Eq.~(\ref{eq:gh_modular}) (with the obvious replacement of $k,l$ by $a,b$).
%Moreover, the lifted relation $h^2=(-1)^F$ implies that spacetime fermions are projected out in perturbation theory; hence the resulting theory is genuinely non-supersymmetric.

\textbf{Untwisted sector (massless bosons)}

Under the action $h:(X^6,X^7,X^8,X^9)\to(X^6,X^8,X^7,-X^9)$ on $T^4$, the remaining $\dim \M^{\text{Het}}_{T^4/h}=41$ massless bosons are one dilaton, six metrics, two B-fields, and $32$ Wilson lines:

\begin{equation}
\begin{aligned}
\phi, (G_{66}+G_{77},G_{88},G_{99},G_{67},G_{68}+G_{78},G_{69}+G_{79}),\\
(B_{68}-B_{78},B_{69}+B_{79}), A^a_1,\cdots,A_4^a,1\leq a\leq \dim(\U(1)^{8}).
\end{aligned}
\end{equation}
In the $T^5$ cases, there are additionally $13$ massless states to match $\dim \M^{\text{Het}}_{T^5/h}=54$. The rank of gauge symmetry is $\dim I_{d-2,d+6}=2d+4$, which includes graviphotons.

\textbf{Twisted sector (tachyon)}

In twisted sectors, basically there are no new massless states, but tachyonic states would arise in some region of the moduli space. The existence of tachyons does not mean the inconsistency of the theory immediately, but implies the wrong choice of vacua \cite{hellerman2008stablevacuumtachyonice8,Kaidi_2021}. 
\newpage

\section{Review of Type II superstrings on a K3 surface (supersymmetric)}\label{sec:K3}

In this section, we review the construction of Type II superstring theories compactified on $T^4/\Z_2$ \cite{Gregori_1998}, which is an orbifold limit of a $K3$ surface. We summarize the bosonic massless spectrum in Tables \ref{tab:IIAspectrum_K3} and \ref{tab:IIBspectrum_K3}.
\begin{comment}
The traces over the Hilbert space of NS/R sector can be written as follows:

\begin{equation}
    \begin{aligned}
        \tr_{\mathrm{NS}}q^{H_{\mathrm{NS}}}=&\frac{1}{\eta^4}\vartheta_{00}^4,\\
        \tr_{\mathrm{NS}}(-1)^{F_L}q^{H_{\mathrm{NS}}}=&\frac{1}{\eta^4}\vartheta_{01}^4,\\
        \tr_{\mathrm{R}}q^{H_{\mathrm{R}}}=&\frac{1}{\eta^4}\vartheta_{10}^4,\\
        \tr_{\mathrm{R}}(-1)^{F_L}q^{H_{\mathrm{R}}}=&\frac{1}{\eta^4}\vartheta_{11}^4,
    \end{aligned}
\end{equation}

\begin{equation}
    \begin{aligned}
        \mathrm{NS}_+(\tau)=&,\\
        \mathrm{NS}_-(\tau)=&,&&\\
        \mathrm{R}_+(\tau)=&,\\
        \mathrm{R}_-(\tau)=&,&&\\
    \end{aligned}
\end{equation}

    \begin{equation}
    \begin{aligned}
        T:&\qty(\mathrm{NS}_+,\mathrm{NS}_-,\mathrm{R}_+,\mathrm{R}_-)\\
        \to&\qty(\mathrm{NS}_-,\mathrm{NS}_+,\mathrm{R}_-,\mathrm{R}_+),\\
S:&\qty(\mathrm{NS}_+,\mathrm{NS}_-,\mathrm{R}_+,\mathrm{R}_-)\\
\to&\qty(\mathrm{NS}_+,\mathrm{R}_+,\mathrm{NS}_-,-i\mathrm{R}_-).
    \end{aligned}
\end{equation}

\begin{equation}
\begin{aligned}
S\cdot\mathrm{NS}_+=&\mathrm{NS}_+,\\
    S^{\pm}\cdot\mathrm{NS}_-=&\mathrm{R}_+,\\
    S^{\pm}\cdot\mathrm{R}_-=&-i\mathrm{R}_-,\\
\end{aligned}
\end{equation}
where $F_L$ is the left-moving fermion number.
\end{comment}

\subsection{Type II superstrings on \texorpdfstring{$T^4$}{T4}}

The Hilbert space of Type II superstrings on $T^4$ is given by
\begin{equation}
\H_{T^4}=\H^{4,4}_{\text{Boson}}\otimes\H^{8,8}_{\text{Fermion}}\otimes\H_{\Gamma_{4,4}},
\end{equation}
with the GSO projection. This is the Hilbert space for the sigma model whose target space is $\R^4\times T^4$ in the light-cone formalism. Here $\H^{4,4}_{\text{Boson}}\otimes \H_{\Gamma_{4,4}}$ is bosonic, while $\H^{8,8}_{\text{Fermion}}$ is fermionic. The indices $4$ or $8$ stand for the dimensions of target space in left/right sector. Also $\Gamma_{4,4}=H^1(T^4;\Z)\oplus H_1(T^4;\Z)$ is an even self-dual momentum lattice for $T^4$.
The torus partition function is given by

\begin{equation}
    \begin{aligned}
        Z^{0,0}\chvec{0}{0}
        =&\frac{1}{(\mathrm{Im}\tau)^2 |\eta|^{8}}\frac{1}{2\eta^4}\sum_{a,b=0}^1(-1)^{a+b+ab}\tchar{a}{b}^4
\frac{1}{2\bar\eta^4}\sum_{\bar a,\bar b=0}^1(-1)^{\bar a+\bar b+\mu\bar a\bar b }\tbchar{\bar a}{\bar b}^4 \Gamma_{4,4}^{0,0}\chvec{0}{0},\\
    \end{aligned}
\end{equation}
where $\mu=0$ for Type IIA and $\mu=1$ for Type IIB. Here $(\vartheta_{00},\vartheta_{01})$ corresponds to the NS sector, while $(\vartheta_{10},\vartheta_{11})$ corresponds to the R sector. The factor $\Gamma_{4,4}^{0,0}\chvec{0}{0}$ comes from $\H_{\Gamma_{4,4}}$ and is given by

\begin{equation}
    \begin{aligned}
     \Gamma_{4,4}^{0,0}\chvec{0}{0}=&\tr_{\H_{\Gamma_{4,4}}}q^{L_0}\bar{q}^{\bar L_0}
     =\frac{1}{\eta^4\bar \eta^4}\sum_{p\in\Gamma_{4,4}}q^{\frac{1}{2}p_L^2}\bar q^{\frac{1}{2}p_R^2}.
    \end{aligned}
\end{equation}

\subsection{Type II superstrings on \texorpdfstring{$T^4/\Z_2$}{T4/Z2} (supersymmetric)}

Then we consider taking the following quotient of $T^4$:

\begin{equation}
    \begin{aligned}
        g :
\begin{cases}
X^6 \rightarrow -X^6,  \\
X^7 \rightarrow -X^7, \\
X^8 \rightarrow -X^8, \\
X^9 \rightarrow -X^9,
\end{cases}\quad  \begin{cases}
(\psi^6,\tilde{\psi}^6) \rightarrow -(\psi^6,\tilde{\psi}^6) ,\\
(\psi^7,\tilde{\psi}^7)  \rightarrow -(\psi^7,\tilde{\psi}^7) ,\\
(\psi^8,\tilde{\psi}^8)  \rightarrow -(\psi^8,\tilde{\psi}^8) ,\\
(\psi^9,\tilde{\psi}^9)  \rightarrow -(\psi^9,\tilde{\psi}^9) ,\\
\end{cases}
    \end{aligned}
\end{equation}
and corresponding orbifold theory: Type II superstring theories on $T^4/\Z_2$. The action of $g$ on $T^4$ has 16 fixed points, and $T^4/\Z_2$ is an orbifold limit of a K3 surface with $16$ singular points. We denote the new Hilbert space as $\H_{K3}$, which is given by 

\begin{equation}
\begin{aligned}
        \H_{K3}=&\frac{1+g}{2}\H_{T^4}\quad\text{(untwisted sector)}\\
        \oplus&\frac{1+g}{2}\H^g_{T^4}\quad\text{($g$-twisted sector)},
\end{aligned}
\end{equation}
where $\H^g_{T^4}$ is the $g$-twisted Hilbert space, which means that the boundary conditions of fields are changed by the action of $g$. For $p=(p_6,p_7,p_8,p_9)\in\Gamma_{4,4}$, $g$ acts as

\begin{equation}
    g\ket{p}=\ket{-p}.
\end{equation}

\begin{comment}
    $g=diag(-1,-1,-1,-1)\in \SO(4)$ can be seen as $g=diag(-1,-1)\times diag(-1,-1)\in \SO(2)\times \SO(2)$  up lifts to $g^2=1 \in \Spin(4)$.
\begin{equation}
    \begin{aligned}
O_8=&O_4^{(\R^4)}O_4^{(T^4)}+V_4^{(\R^4)}V_4^{(T^4)},\\
        V_8=&O_4^{(\R^4)} V_4^{(T^4)}++V_4^{(\R^4)}O_4^{(T^4)},\\
S_8=&S_4^{(\R^4)}S_4^{(T^4)}++C_4^{(\R^4)}C_4^{(T^4)},\\
C_8=&S_4^{(\R^4)}C_4^{(T^4)}++C_4^{(\R^4)}S_4^{(T^4)},\\
    \end{aligned}
\end{equation}

$g$ also acts characters as follows:

\begin{equation}
\begin{aligned}
        g:&(O_4,V_4,S_4,C_4)_{\R^4}\to (O_4,V_4,S_4,C_4)_{\R^4},\\
        &(O_4,V_4,S_4,C_4)_{T^4}\to (O_4,-V_4,S_4,-C_4)_{T^4}
\end{aligned}    
\end{equation}
\end{comment}

Now we can define the partition function and its building blocks of the new theory as follows:

\begin{equation}
\begin{aligned}
        Z=&\frac{1}{2}\sum_{c,d=0}^1 Z^{0,0}\chvec{c}{d},\\
        Z^{0,0}\chvec{c}{d}=&\tr_{\H^{g^c}}g^d \q.
\end{aligned}    
\end{equation}
We refer to $Z^{0,0}\chvec{0}{0}$ and $Z^{0,0}\chvec{0}{1}$ as the \textit{untwisted sector} of Type II theories on $T^4/\Z_2$:
\begin{equation}
\begin{aligned}
Z^{0,0}\chvec{0}{1}
=&\tr_{\H_{T^4}}g \q\\
=&\frac{1}{(\Im\tau)^2|\eta|^{8}}
\frac{1}{2\eta^2}\sum_{a,b=0}^1(-1)^{a+b+ab}\tchar{a}{b}^2
\frac{1}{2\bar\eta^2}\sum_{\bar a,\bar b=0}^1(-1)^{\bar a+\bar b+\mu\bar a\bar b }\tbchar{\bar a}{\bar b}^2\\
&\quad\times
\frac{1}{|\eta|^4}
\tchar{a}{b+1}\tchar{a}{b-1}\tbchar{\bar a}{\bar b+1}\tbchar{\bar a}{\bar b-1}
\Gamma_{4,4}^{0,0}\chvec{0}{1}\\
\end{aligned}
\end{equation}
Using the modular transformation, we define an additional sector as follows:

\begin{equation}
\begin{aligned}
        Z^{0,0}\chvec{1}{0}=&S\cdot Z^{0,0}\chvec{0}{1},\\
        Z^{0,0}\chvec{1}{1}=&T\cdot Z^{0,0}\chvec{1}{0}.
\end{aligned}
\end{equation}
We refer to $Z^{0,0}\chvec{1}{0}$ and $Z^{0,0}\chvec{1}{1}$ as the \textit{twisted sector} of Type II theories on $T^4/\Z_2$. Then the result is

\begin{equation}
\begin{aligned}
Z
&=\frac{1}{(\Im\tau)^2|\eta|^{8}}
\frac{1}{2\eta^2}\sum_{a,b=0}^1(-1)^{a+b+ab}\tchar{a}{b}^2
\frac{1}{2\bar\eta^2}\sum_{\bar a,\bar b=0}^1(-1)^{\bar a+\bar b+\mu\bar a\bar b }\tbchar{\bar a}{\bar b}^2\\
&\quad\times
\frac{1}{2|\eta|^4}\sum_{c,d=0}^1
\tchar{a+c}{b+d}
\tchar{a-c}{b-d}
\tbchar{\bar a+c}{\bar b+d}
\tbchar{\bar a-c}{\bar b-d}
\Gamma_{4,4}^{0,0}\chvec{c}{d},
\end{aligned}
\end{equation}
where
\begin{equation}
\begin{aligned}
    \Gamma_{4,4}^{0,0}\chvec{0}{0}=&\frac{1}{\eta^4\bar \eta^4}\sum_{p\in\Gamma_{4,4}}q^{\frac{1}{2}p_L^2}\bar q^{\frac{1}{2}p_R^2},\\
    \quad\Gamma_{4,4}^{0,0}\chvec{c}{d}=&\frac{16|\eta|^{4}}{|\tchar{1+c}{1+d}|^4}, (c,d)\neq (0,0).
\end{aligned}
\end{equation}
The total partition function is modular invariant.

\begin{figure}
    \centering
    \begin{tikzpicture}[>=stealth, baseline=(current bounding box.center)]

% Nodes
\node (A) at (0,2) {$Z^{0,0}\chvec{0}{0}$};
\node (B) at (3,2) {$Z^{0,0}\chvec{0}{1}$};
\node (C) at (0,0) {$Z^{0,0}\chvec{1}{0}$};
\node (D) at (3,0) {$Z^{0,0}\chvec{1}{1}$};

% Arrows
\draw[->] (A) edge[loop above] node[above] {\small $T, S$} (A);
\draw[->] (B) edge[loop above] node[above] {\small $T$} (B);
\draw[->] (D) edge[loop right] node[right] {\small $S$} (D);

\draw[->] (A) -- (B) node[midway, above] {};
\draw[->] (C) -- (D) node[midway, below] {\small $T$};
\draw[->] (D) -- (C) node[midway, below] {\small $T$};

\draw[->] (B) -- (C) node[midway, sloped, above] {\small $S$};
\draw[->] (C) -- (B) node[midway, sloped, below]{} ;

\end{tikzpicture}
    \caption{Modular transformations of partition function on $T^4/\Z_2$.}
\end{figure}
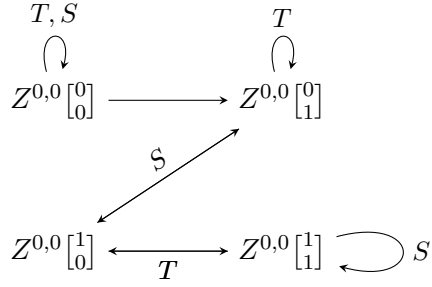

\subsection{Spectrum and moduli space}

\subsubsection{IIA on K3}

The massless spectrum of type IIA theory on $T^4/\Z_2$ contains the $6d ~\N=(1,1)$ supergravity multiplet and the  $20 ~U(1)$ vector multiplets (Table \ref{tab:IIAspectrum_K3}).

\begin{table}[h]
\centering
\renewcommand{\arraystretch}{1.3}
\begin{tabular}{c|c|c!{\vrule width 0.4pt}c|c|}
\multicolumn{2}{c|}{} &
\multicolumn{2}{c|}{untwisted sector $\qty(\frac{1+g}{2}\H_{T^4})$} &
$g$-twisted sector $\qty(\frac{1+g}{2}\H^g_{T^4})$ \\
\hline
\multicolumn{2}{c|}{$\N=(1,1)$ multiplets} &
SUGRA & $U(1)^4$ Vector & $U(1)^{16}$ Vector \\
\hline
\multicolumn{2}{c|}{NSNS components} &
$G_{\mu\nu},B_{\mu\nu},\phi$  & $16$ scalars & $64$ scalars \\
\hline
\multicolumn{2}{c|}{RR components} &
$C_{\mu},C^+_{\mu ab}$& $C^-_{\mu ab},C_{\mu\nu\rho}$ & $16$ vectors \\
\hline
\end{tabular}\caption{Bosonic massless spectrum of IIA on \texorpdfstring{$T^4/\Z_2$}{T4/Z2}. $\mu,\nu,\rho$ are indices on $\R^4$, while  $a,b=6,7,8,9,$ and $\pm$ denotes the SD/ASD conditions on indices $a,b$. There are $81$ scalar fields and 24 $U(1)$ vectors.}\label{tab:IIAspectrum_K3}
\end{table}

The moduli space is locally given by \cite{Aspinwall:1994rg}:

\begin{equation}
\begin{aligned}
    \mathcal{M}^{\text{IIA}}_{\text{K3}}=&\frac{O(4,20)}{O(4)\times O(20)}\times\R_{\text{dilaton}},
\end{aligned}
\end{equation}
which is $81$-dimensional. 17 components come from the untwisted sector $\frac{1+g}{2}\H_{T^4}$, while 64 components come from the twisted sector  $\frac{1+g}{2}\H^g_{T^4}$.

\subsubsection{IIB on K3}
The massless spectrum of Type IIB theory on $T^4/\Z_2$ contains the $6d ~\N=(2,0)$ supergravity multiplet and the $21~U(1)$ tensor multiplets (Table \ref{tab:IIBspectrum_K3}).

\begin{table}[h]
\centering
\renewcommand{\arraystretch}{1.3}
\begin{tabular}{c|c|c!{\vrule width 0.4pt}c|c|}
\multicolumn{2}{c|}{} &
\multicolumn{2}{c|}{untwisted sector $\qty(\frac{1+g}{2}\H_{T^4})$} &
$g$-twisted sector $\qty(\frac{1+g}{2}\H^g_{T^4})$ \\
\hline
\multicolumn{2}{c|}{$\N=(2,0)$ multiplets}&
SUGRA & $5$ Tensors
&  $16$ Tensors \\
\hline
\multicolumn{2}{c|}{NSNS components}
& $(G_{\mu\nu},B^{+}_{\mu\nu})$& $(B_{\mu\nu}^{-},\phi,G_{ab},B_{ab})$
& $64$ scalars \\ 
\hline
\multicolumn{2}{c|}{RR components}
& $C^{+}_{\mu\nu},C^{+}_{\mu\nu ab}$ & $C,C_{ab},C_{6789},C^{-}_{\mu \nu},C^{-}_{\mu\nu ab}$
& $16$ scalars, $16$ ASD  tensors\\
\hline
\end{tabular}\caption{Bosonic massless spectrum of IIB on $T^4/\Z_2$. $\mu,\nu,\rho$ are indices on $\R^4$, while  $a,b=6,7,8,9.$ $\pm$ denotes SD/ASD on the indices $a,b$. There are $105$ scalar fields and 26 $U(1)$ tensor fields.}\label{tab:IIBspectrum_K3}
\end{table}

The moduli space is locally \cite{aspinwall1999k3surfacesstringduality}:

\begin{equation}
\begin{aligned}
     \mathcal{M}^{\text{IIB}}_{\text{K3}}=\frac{O(5,21)}{O(5)\times O(21)},
\end{aligned}
\end{equation}
and $\dim \mathcal{M}^{\text{IIB}}_{\text{K3}}=105$.
25 components come from the untwisted sector $\frac{1+g}{2}\H_{T^4}$, while 80 components come from the twisted sector  $\frac{1+g}{2}\H^g_{T^4}$. $26$ SD/ASD tensor fields give $U(1)^{26}$ gauge symmetry.
\subsection{Supergravity}
Away from special points in the moduli space, an explicit worldsheet CFT description of K3 compactifications is generally not available. Nevertheless, the low-energy supergravity description remains well defined and is a convenient framework for identifying the moduli space and the massless field content. 

The moduli space of Ricci-flat metrics on K3 surfaces is given by \cite{Itoh1994}:

\begin{equation}
    \R_{\text{Volume}}\times \frac{O(3,19)}{O(3)\times O(19)},
\end{equation}
which has $58$ dimensions. Additionally, the second homology class of K3 surfaces is:
\begin{equation}
 H_2(K3;\Z)=\Gamma_{3,19},
\end{equation}
which is $22$-dimensional even self-dual lattice. Both in Type IIA and Type IIB, there are $81=58+22+1$ massless scalar fields coming from metrics, B-field, and dilaton. In Type IIA theory, the moduli space is given by

\begin{equation}
    \M^{\N=(1,1) \text{ SUGRA}}_{K3}= \R_{\text{Volume}}\times \frac{O(3,19)}{O(3)\times O(19)}\times \R^{3,19}_{B}\times \R_{\text{dilaton}}.
\end{equation}

In Type IIB theory, there are additional $24$ massless fields coming from RR fields $C_0,C_2,C_4$ wrapping homology cycles of a K3 surface $H_0\oplus H_2\oplus H_4=\Gamma_{4,20}$. Then the moduli space is

\begin{equation}
    \M^{\N=(2,0) \text{ SUGRA}}_{K3}= \R_{\text{Volume}}\times \frac{O(3,19)}{O(3)\times O(19)}\times \R^{3,19}_{B}\times\R^{4,20}_{RR}\times \R_{\text{dilaton}},
\end{equation}
which is $105$-dimensional.

\newpage
\section{Type II orbifold (non-supersymmetric)}\label{sec:Enr}
In this section, we construct Type II superstring theories compactified on $(T^4/\Z_2)/\Z_2$, which is an orbifold limit of an Enriques surface. We summarize massless and tachyonic spectrum in Tables \ref{table:IIAspectrum_Enr} and \ref{table:IIBspectrum_Enr}.

\subsection{A fixed-point-free involution \texorpdfstring{$h$}{h} on a K3 surface}

We consider the following quotient of $K3=T^4/\Z_2$:

\begin{equation}\label{eq:hdef}
    \begin{aligned}
        h :
\begin{cases}
X^6 \rightarrow -X^6,\\
X^7 \rightarrow -X^7 + \pi R_7 ,\\
X^8 \rightarrow X^8 + \pi R_8,\\
X^9 \rightarrow X^9. \\
\end{cases}
    \end{aligned}
\end{equation}
It should be noted that $h$ has no fixed points (since $X^8\to X^8+\pi R_8$). It can be easily seen that $h$ maps a $g$-fixed point to another $g$- fixed point, which means that $h$ reduced the number of singular points $16$ to $8$. Geometrically, the quotient of $K3=T^4/\Z_2$ by $h$ is called an \textit{Enriques surface} (more precisely its orbifold limit). While this involution and possibility of new orbifold were already mentioned elsewhere \cite{Acharya:2019mcu,Gopakumar_1996,Braun_2010,Dabholkar_1996,Acharya:2022shu}, no explicit calculation has been presented yet.

\subsection{Orbifold}

The action of $h$ on the worldsheet operators are:
\begin{equation}
    \begin{aligned}
        h :
\begin{cases}
X^6 \rightarrow -X^6,\\
X^7 \rightarrow -X^7 + \pi R_7 ,\\
X^8 \rightarrow X^8 + \pi R_8,\\
X^9 \rightarrow X^9. \\
\end{cases}
\quad\begin{cases}
(\psi^6,\tilde{\psi}^6) \rightarrow -(\psi^6,\tilde{\psi}^6) ,\\
(\psi^7,\tilde{\psi}^7)  \rightarrow -(\psi^7,\tilde{\psi}^7) ,\\
(\psi^8,\tilde{\psi}^8)  \rightarrow (\psi^8,\tilde{\psi}^8) ,\\
(\psi^9,\tilde{\psi}^9)  \rightarrow (\psi^9,\tilde{\psi}^9) .\\
\end{cases}
    \end{aligned}
\end{equation}
Next, we should specify the action of $h$ on the Hilbert space. Up to half shifts, $h$ can be viewed as an element of  $\SO(8)$:

\begin{equation}
    h\simeq\qty( \begin{matrix}
        1_{4}\\
        &-1_2\\
        &&1_{2}\\
    \end{matrix})\in\SO(8).
\end{equation}
Since $h$ is a $\pi$ rotation on $\R^8$, the relation $h^2=1\in\SO(8)$ lifts non-trivially to $h^2=-1\in\Spin(8)$. Then, we can view $h^2$ as $(-1)^F$, where $F$ is the spacetime fermion number. (In contrast, $g=(1_4,-1_4)$ can be viewed as $\text{diag}(1_4,-1_2,1_2)\times\text{diag}(1_4,1_2,-1_2 )\in \SO(8)$, so the relation $g^2=1$ lifted trivially to $g^2=(-1)\times (-1)=1\in\Spin(8)$.)

For $p=(p_6,p_7,p_8,p_9)\in\Gamma_{4,4}$, $h$ acts

\begin{equation}
\begin{aligned}
        h\ket{p_6,p_7,p_8,p_9}=&(-1)^{R_7p_7+R_8p_8}\ket{-p_6,-p_7,p_8,p_9},\\
        hg\ket{p_6,p_7,p_8,p_9}=&(-1)^{R_7p_7+R_8p_8}\ket{p_6,p_7,-p_8,-p_9}.
\end{aligned}
\end{equation}

We define the Hilbert space of new orbifolding theory and call it  $\H_{\text{Enriques}}$:

\begin{equation}\label{eq:HilEnr}
    \begin{aligned}
        \H_{\text{Enriques}}=&\frac{1+h+h^2+h^3}{4}\H_{K3} \quad\text{(untwisted sector)}\\
        \oplus&\frac{1+h+h^2+h^3}{4}\H^h_{K3}\quad\text{($h$-twisted sector)}\\
        \oplus&\frac{1+h+h^2+h^3}{4}\H^{h^2}_{K3}\quad\text{($h^2$-twisted sector)}\\
        \oplus&\frac{1+h+h^2+h^3}{4}\H^{h^3}_{K3}\quad\text{($h^3$-twisted sector)}.\\
    \end{aligned}
\end{equation}

We define the partition function and its building blocks for the new theory as follows:
\begin{equation}
\begin{aligned}
    Z=&\frac{1}{4}\sum_{k,l=0}^3 Z^{k,l},\quad Z^{k,l}=\sum_{c,d=0}^1 Z^{k,l}\chvec{c}{d},\\
    &Z^{k,l}\chvec{c}{d}=\tr_{\H_{T^4}^{g^c h^k}}g^dh^l\q,
\end{aligned} 
\end{equation}
where $\H_{T^4}^{g^c h^k}$ is $g^ch^k$-twisted Hilbert space. These functions can be computed in two different ways: (1) as a trace over the Hilbert space $\H^{g^c h^k}$, or (2) certain modular transformation from other functions computed earlier:

\begin{equation}\label{eq:gh_modular}
    \begin{aligned}
        T\cdot Z^{k,l}\chvec{c}{d}=&Z^{k,k+l}\chvec{c}{c+d},\\
        S\cdot Z^{k,l}\chvec{c}{d}=&Z^{l,-k}\chvec{d}{-c}.
    \end{aligned}
\end{equation}
As a result, $Z^{k,l}\chvec{c}{d}$ is explicitly given as follows:

\begin{equation}
\begin{aligned}
Z^{k,l}\chvec{c}{d}=&\tr_{\H^{g^ch^k}}g^dh^l\q\\
=&\frac{1}{(\Im\tau)^2|\eta|^{8}}
\frac{1}{2\eta^2}\sum_{a,b=0}^1(-1)^{a+b+ab}\tchar{a}{b}^2
\frac{1}{2\bar\eta^2}\sum_{\bar a,\bar b=0}^1(-1)^{\bar a+\bar b+\mu\bar a\bar b }\tbchar{\bar a}{\bar b}^2\\
&\quad\times
\frac{1}{|\eta|^4}e^{-\frac{\pi i}{2}k(b-\bar b)}
\tchar{a+c+k}{b+d+l}
\tchar{a-c}{b-d}
\tbchar{\bar a+c+k}{\bar b+d+l}
\tbchar{\bar a-c}{\bar b-d}
\Gamma_{4,4}^{k,l}\chvec{c}{d} \\
%=&\frac{1}{(\Im\tau)^2|\eta|^{8}}\cdot \frac{1}{4\eta^4\bar\eta^4}\left|\sum_{a,b=0}^1(-1)^{a+b}e^{-\frac{\pi i}{2}kb}\tchar{a}{b}^2\tchar{a+c+k}{b+d+l}\tchar{a-c}{b-d}\right|^2\cdot\Gamma_{4,4}^{k,l}\chvec{c}{d},\\
\end{aligned}
\end{equation}
where

\begin{equation}
    \Gamma^{a,b}_{4,4}\chvec{c}{d}=\tr_{\H_{\Gamma_{4,4}}^{h^ag^c}}h^b g^dq^{L_0}\bar{q}^{\bar L_0}.
\end{equation}

Computations of $\Gamma^{a,b}_{4,4}\chvec{c}{d}$ are contained in Appendix  \ref{sec:Twisted bosonic fock space}.
Because these building blocks satisfy modular transformation rules (equation \eqref{eq:gh_modular}), the total partition function $Z$ is modular invariant.

\textbf{No fermions and no supersymmetry}

Perturbative fermions are projected under the action of $h^2=(-1)^F$:
\begin{equation}
    \begin{aligned}
        \frac{1+h+h^2+h^3}{4}=\frac{1+h}{2}\frac{1+(-1)^F}{2}.
    \end{aligned}
\end{equation}
This is consistent with the fact that Enriques surface does not admit a spin structure. It follows from $b_2^+-b_2^-=-8\notin 16\Z $ and Rokhlin's theorem. (But D-branes may give rise to charged massless fermions at strong coupling). Consequently, there are no supersymmetry.

This fact simplifies the partition function. The insertion of $\frac{1+(-1)^F}{2}$ corresponds to the condition $\bar a=a$, and the insertion of $\frac{1+h}{2}$ corresponds to the sum of $l=0,1$:
\begin{equation}
    \begin{aligned}
        \frac{1}{4}\sum_{l=0}^3Z^{k,l}\chvec{c}{d}=&\frac{1}{(\Im\tau)^2|\eta|^{8}}
\frac{1}{2\eta^2}\sum_{a,b,\bar b=0}^1(-1)^{b+ab}\tchar{a}{b}^2
\frac{1}{2\bar\eta^2}(-1)^{\bar b+\mu a\bar b }\tbchar{ a}{\bar b}^2\\
\quad\times
\frac{1}{|\eta|^4}&\frac{1}{2}\sum_{l=0}^1 e^{-\frac{\pi i}{2}k(b-\bar b)}
\tchar{a+c+k}{b+d+l}
\tchar{a-c}{b-d}
\tbchar{a+c+k}{\bar b+d+l}
\tbchar{a-c}{\bar b-d}
\Gamma_{4,4}^{k,l}\chvec{c}{d}\\
=&\frac{1}{8(\Im\tau)^2\eta^8\bar \eta^8}
\sum_{a,l=0}^1 
\left|\sum_{b=0}^1(-1)^{b}e^{-\frac{\pi i}{2}kb}\tchar{a}{b}^2\tchar{a+c+k}{b+d+l}
\tchar{a-c}{b-d}\right|^2
\Gamma_{4,4}^{k,l}\chvec{c}{d}.
    \end{aligned}
\end{equation}
\subsection{Untwisted sector}

The Hilbert space of the untwisted sector is given by
\begin{equation}
    \begin{aligned}
&\frac{1+h+h^2+h^3}{4}\H_{K3}\\
        =&\frac{1+h+h^2+h^3}{4}\frac{1+g}{2}\qty(\H_{T^4}\oplus\H^{g}_{T^4}).
    \end{aligned}
\end{equation}
Its partition function is 
\begin{equation}\label{eq:untwisted_trace}
    \begin{aligned}
&\tr_{\H_{T^4}\oplus\H_{T^4}^{g}}\frac{1+h+h^2+h^3}{4}\frac{1+g}{2}\q
        =\frac{1}{2}\sum_{c,d,l=0}^1 Z^{0,l}\chvec{c}{d}\\
        =&\frac{1}{16(\Im\tau)^2\eta^8\bar \eta^8}
\sum_{a,c,d,l=0}^1 
\left|\sum_{b=0}^1(-1)^{b}\tchar{a}{b}^2\tchar{a+c}{b+d+l}
\tchar{a-c}{b-d}\right|^2
\Gamma_{4,4}^{0,l}\chvec{c}{d} \\
=&\frac{1}{16(\Im\tau)^2\eta^8\bar \eta^8}
\sum_{a,c,d=0}^1 
\left|\sum_{b=0}^1(-1)^{b}\tchar{a}{b}^2\tchar{a+c}{b+d}
\tchar{a-c}{b-d}\right|^2
\Gamma_{4,4}^{0,0}\chvec{c}{d}\quad (l=0) \\
&\quad +\frac{1}{16(\Im\tau)^2\eta^8\bar \eta^8}
\sum_{a,c,d=0}^1 
\left|\sum_{b=0}^1(-1)^{b}\tchar{a}{b}^2\tchar{a}{b+d+1}
\tchar{a}{b-d}\right|^2
\Gamma_{4,4}^{0,1}\chvec{0}{d}.\quad (l=1, \text{then } c=0). 
    \end{aligned}
\end{equation}

\subsubsection{untwisted NSNS sector}

We take $a=0$ in equation \eqref{eq:untwisted_trace}:

\begin{equation}
    \begin{aligned}
        &\frac{1}{16(\Im\tau)^2\eta^8\bar \eta^8}
\sum_{c,d=0}^1 
\left|\sum_{b=0}^1(-1)^{b}\tchar{0}{b}^2\tchar{c}{b+d}
\tchar{-c}{b-d}\right|^2
\Gamma_{4,4}^{0,0}\chvec{c}{d}\quad (l=0) \\
&\quad+\frac{1}{16(\Im\tau)^2\eta^8\bar \eta^8}
\sum_{d=0}^1 
\left|\sum_{b=0}^1(-1)^{b}\tchar{0}{b}^2\tchar{0}{b+d+1}
\tchar{0}{b-d}\right|^2
\Gamma_{4,4}^{0,1}\chvec{0}{d}\quad (l=1, \text{then } c=0)\\
        =&\frac{1}{16(\Im\tau)^2\eta^8\bar \eta^8}
\left|\sum_{b=0}^1(-1)^{b}\tchar{0}{b}^4\right|^2
\Gamma_{4,4}^{0,0}\chvec{0}{0} \quad (c=0,\text{then } d=0)\\
+&\frac{1}{16(\Im\tau)^2\eta^8\bar \eta^8}
\sum_{d=0}^1\left|(-1)^{d}\tchar{0}{d}^2\tchar{1}{2d}
\tchar{1}{0}\right|^2
\Gamma_{4,4}^{0,0}\chvec{1}{d}\quad(c=1, \text{then } b=d)\\
&\quad+\frac{1}{16(\Im\tau)^2\eta^8\bar \eta^8}
\left|\sum_{b=0}^1(-1)^{b}\tchar{0}{b}^2\tchar{0}{b+1}
\tchar{0}{b}\right|^2
\Gamma_{4,4}^{0,1}\chvec{0}{0}\quad(d=0)\\
&\quad +\frac{1}{16(\Im\tau)^2\eta^8\bar \eta^8} 
\left|\sum_{b=0}^1(-1)^{b}\tchar{0}{b}^2\tchar{0}{b}
\tchar{0}{b-1}\right|^2
\Gamma_{4,4}^{0,1}\chvec{0}{1}\quad (d=1)\\
=&\frac{1}{(\Im \tau)^2}\qty(56+...).
    \end{aligned}
\end{equation}
These 56 massless NS–NS states decompose (in the light-cone description) into 9 graviton, 6 B-fields, and 41 scalar fields.
Here we used
\begin{equation}
\begin{aligned}
        \Gamma^{0,1}_{4,4}\chvec{0}{0}
        =&\frac{1}{\eta^4 \bar{\eta}^4}+...,\\
        \Gamma^{0,1}_{4,4}\chvec{0}{1}
        =&\frac{1}{\eta^4 \bar{\eta}^4}+....
    \end{aligned}
\end{equation}

\subsubsection{untwisted RR sector}

We take $a=1$ (then $b=0$ since $\vartheta_{11}=0$) in equation \eqref{eq:untwisted_trace}:
\begin{equation}
    \begin{aligned}
        &\frac{1}{16(\Im\tau)^2\eta^8\bar \eta^8}\sum_{c,d=0}^1
\left|\tchar{1}{0}^2\tchar{1+c}{d}
\tchar{1-c}{-d}\right|^2
\Gamma_{4,4}^{0,0}\chvec{c}{d}\\
&\quad +\frac{1}{16(\Im\tau)^2\eta^8\bar \eta^8} 
\sum_{c,d=0}^1\left|\tchar{1}{0}^2\cancel{\tchar{1}{d+1}
\tchar{1}{-d}}\right|^2
\Gamma_{4,4}^{0,1}\chvec{0}{d}\quad \\
=&\frac{1}{16(\Im\tau)^2\eta^8\bar \eta^8}
\left|\tchar{1}{0}^4\right|^2
\Gamma_{4,4}^{0,0}\chvec{0}{0}\quad (c=0, \text{then } d=0)\\
&\quad+\frac{1}{16(\Im\tau)^2\eta^8\bar \eta^8}
\sum_{d=0}^1\left|\tchar{1}{0}^2\tchar{0}{d}^2\right|^2
\Gamma_{4,4}^{0,0}\chvec{1}{d}\quad (c=1)\\
=&\frac{1}{(\Im\tau)^2}\qty(48+...).
    \end{aligned}
\end{equation}

These $48$ massless RR states can be seen as $12$ U(1) vectors on the IIA side, or $12$ scalar fields and $12$ self-dual/anti-self-dual two-forms on the IIB side. As we will see in subsection \ref{subsec:type0_gravity}, the number $12$ corresponds to the dimension of the homology lattice $H_0\oplus H_2\oplus H_4$ (or Euler number) of Enriques surfaces.

\subsection{\texorpdfstring{$h$}{h}-twisted sector}

The Hilbert space of $h$-twisted sector is 
\begin{equation}
    \begin{aligned}
&\frac{1+h+h^2+h^3}{4}\H^h_{K3}\\
        =&\frac{1+h+h^2+h^3}{4}\frac{1+g}{2}\qty(\H^h_{T^4}\oplus\H^{gh}_{T^4}).
    \end{aligned}
\end{equation}
The partition function is given by

\begin{equation}\label{eq:t_twisted_trace}
    \begin{aligned}
&\tr_{\H_{T^4}^{h}\oplus\H_{T^4}^{gh}}\frac{1+h+h^2+h^3}{4}\frac{1+g}{2}\q
        =\frac{1}{2}\sum_{c,d,l=0}^1 Z^{1,l}\chvec{c}{d}\\
        =&\frac{1}{16(\Im\tau)^2\eta^8\bar \eta^8}
\sum_{a,c,d,l=0}^1 
\left|\sum_{b=0}^1e^{\frac{\pi i}{2}b}\tchar{a}{b}^2\tchar{a+c+1}{b+d+l}
\tchar{a-c}{b-d}\right|^2
\Gamma_{4,4}^{1,l}\chvec{c}{d}\\
=&\frac{1}{16(\Im\tau)^2\eta^8\bar \eta^8}
\sum_{a,c=0}^1 
\left|\sum_{b=0}^1e^{\frac{\pi i}{2}b}\tchar{a}{b}^2\tchar{a+c+1}{b}
\tchar{a-c}{b}\right|^2
\Gamma_{4,4}^{1,0}\chvec{c}{0}\quad (l=0, \text{then } d=0)\\
&\quad+\frac{1}{16(\Im\tau)^2\eta^8\bar \eta^8}
\sum_{a,c=0}^1 
\left|\sum_{b=0}^1e^{\frac{\pi i}{2}b}\tchar{a}{b}^2\tchar{a+c+1}{b+c+1}
\tchar{a-c}{b-c}\right|^2
\Gamma_{4,4}^{1,1}\chvec{c}{c},\quad (l=1, \text{then } d=c)
    \end{aligned}
\end{equation}
here we used

\begin{equation}
    \begin{aligned}
        \Gamma^{1,0}\chvec{0}{1}=\Gamma^{1,0}\chvec{1}{1}=0,\\
        \Gamma^{1,1}\chvec{0}{1}=\Gamma^{1,1}\chvec{1}{0}=0.
    \end{aligned}
\end{equation}

\subsubsection{\texorpdfstring{$h$}{h}-twisted NSNS sector}
We take $a=0$ in equation \eqref{eq:t_twisted_trace}:

\begin{equation}
    \begin{aligned}
        &\frac{1}{16(\Im\tau)^2\eta^8\bar \eta^8}
\sum_{c=0}^1 
\left|\tchar{0}{0}^2\tchar{c+1}{0}
\tchar{c}{0}\right|^2
\Gamma_{4,4}^{1,0}\chvec{c}{0}\quad (b=0)\\
+&\frac{1}{16(\Im\tau)^2\eta^8\bar \eta^8}
\sum_{c=0}^1 
\left|\tchar{0}{1}^2\tchar{c+1}{c+2}
\tchar{c}{1-c}\right|^2
\Gamma_{4,4}^{1,1}\chvec{c}{c}\quad (b=1)\\
=&\frac{1}{(\Im\tau)^2}\qty(2q^{\frac{1}{8}(R_8^2-4)}\bar q^{\frac{1}{8}(R_8^2-4)}+2q^{\frac{1}{8}(R_7^2-4)}\bar q^{\frac{1}{8}(R_7^2-4)})+....
    \end{aligned}
\end{equation}
These four states are moduli-dependent tachyons.

\subsubsection{\texorpdfstring{$h$}{h}-twisted RR sector} 

We take $a=1$ (then $b=0$) in equation \eqref{eq:t_twisted_trace}:

\begin{equation}
    \begin{aligned}
        =&\frac{1}{16(\Im \tau)^2\eta^8\bar \eta^8}
\sum_{c=0}^1 
\left|\tchar{1}{0}^2\tchar{c}{0}
\tchar{1-c}{0}\right|^2
\Gamma_{4,4}^{1,0}\chvec{c}{0}\\
&\quad +\frac{1}{16(\Im\tau)^2\eta^8\bar \eta^8}
\sum_{c=0}^1 
\left|\tchar{1}{0}^2\tchar{c}{c+1}
\tchar{1-c}{-c}\right|^2
\Gamma_{4,4}^{1,1}\chvec{c}{c}\\
=&32q^{\frac{1}{8}R_8^2}\bar q^{\frac{1}{8}R_8^2}+32q^{\frac{1}{8}R_7^2}\bar q^{\frac{1}{8}R_7^2}+...
    \end{aligned}
\end{equation}
The condition $b=0$  comes from $\vartheta_{11}=0$. Here we used:

\begin{equation}
    \begin{aligned}
        \Gamma_{4,4}^{1,0}\chvec{0}{0}
        =&\frac{4}{\eta \bar\eta}q^{\frac{1}{8}R_8^2}\bar q^{\frac{1}{8}R_8^2}+...,\\
            \Gamma_{4,4}^{1,0}\chvec{1}{0}
        =&\frac{4}{\eta \bar\eta}q^{\frac{1}{8}R_7^2}\bar q^{\frac{1}{8}R_7^2}+...,\\
        \Gamma_{4,4}^{1,1}\chvec{0}{0}
        =&\frac{4}{\eta \bar\eta}q^{\frac{1}{8}R_8^2}\bar q^{\frac{1}{8}R_8^2}+...,\\
 \Gamma_{4,4}^{1,1}\chvec{1}{1}
        =&\frac{4}{\eta \bar\eta}q^{\frac{1}{8}R_7^2}\bar q^{\frac{1}{8}R_7^2}+....
    \end{aligned}
\end{equation}

\subsection{\texorpdfstring{$h^2$}{h2}-twisted sector}

The Hilbert space of $h^2$-twisted sector is

\begin{equation}
    \begin{aligned}
&\frac{1+h+h^2+h^3}{4}\H^{h^2}_{K3}\\
        =&\frac{1+h+h^2+h^3}{4}\frac{1+g}{2}\qty(\H^{h^2}_{T^4}\oplus\H^{gh^2}_{T^4}).
    \end{aligned}
\end{equation}
Its partition function is given by

\begin{equation}\label{eq:h^2_twisted_trace}
    \begin{aligned}
&\tr_{\H_{T^4}^{h^2}\oplus\H_{T^4}^{gh^2}}\frac{1+h+h^2+h^3}{4}\frac{1+g}{2}\q
        =\frac{1}{2}\sum_{c,d,l=0}^1 Z^{2,l}\chvec{c}{d}\\
      =&\frac{1}{16(\Im\tau)^2\eta^8\bar \eta^8}
\sum_{a,c,d,l=0}^1 
\left|\sum_{b=0}^1\tchar{a}{b}^2\tchar{a+c}{b+d+l}
\tchar{a-c}{b-d}\right|^2
\Gamma_{4,4}^{0,l}\chvec{c}{d} \\
=&\frac{1}{16(\Im\tau)^2\eta^8\bar \eta^8}
\sum_{a,d,l=0}^1 
\left|\sum_{b=0}^1\tchar{a}{b}^2\tchar{a}{b+d+l}
\tchar{a}{b-d}\right|^2
\Gamma_{4,4}^{0,l}\chvec{0}{d}\quad (c=0)\\
&\quad +\frac{1}{16(\Im\tau)^2\eta^8\bar \eta^8}
\sum_{a,d=0}^1 
\left|\sum_{b=0}^1\tchar{a}{b}^2\tchar{a+1}{b+d}
\tchar{a-1}{b-d}\right|^2
\Gamma_{4,4}^{0,0}\chvec{1}{d} \quad (c=1,\text{then } l=0),
    \end{aligned}
\end{equation}
here we used $\vartheta\chvec{1}{1}=0,\vartheta\chvec{2}{d+l}=\vartheta\chvec{0}{d+l}$ and $\Gamma^{0,1}\chvec{1}{d}=0$.

\subsubsection{\texorpdfstring{$h^2$}{h2}-twisted NSNS sector}
We take $a=0$ in equation \eqref{eq:h^2_twisted_trace}:

\begin{equation}
    \begin{aligned}
        &\frac{1}{16(\Im\tau)^2\eta^8\bar \eta^8}
\sum_{d,l=0}^1 
\left|\sum_{b=0}^1\tchar{0}{b}^2\tchar{0}{b+d+l}
\tchar{0}{b-d}\right|^2
\Gamma_{4,4}^{0,l}\chvec{0}{d}\\
&\quad +\frac{1}{16(\Im\tau)^2\eta^8\bar \eta^8}
\sum_{d=0}^1\left|\tchar{0}{d}^2\tchar{1}{2d}
\tchar{-1}{0}\right|^2
\Gamma_{4,4}^{0,0}\chvec{1}{d} \quad(b=d)\\
=&\frac{1}{(\Im \tau)^2}\qty(q^{-\frac{1}{2}}\bar q^{-\frac{1}{2}}+32+...)
    \end{aligned}
\end{equation}
There is a moduli-independent tachyon and $32$ massless scalars. As we will discuss later, this tachyon would become massive at strong coupling \cite{Bergman:1999km}, and the $32$ massless states would also become massive at strong coupling. Here we used:

\begin{equation}
    \begin{aligned}
     \Gamma_{4,4}^{0,0}\chvec{0}{0}=&q^{-\frac{1}{6}}\bar q^{-\frac{1}{6}}+...,\quad 
     &&\Gamma_{4,4}^{0,0}\chvec{0}{1}=q^{-\frac{1}{6}}\bar q^{-\frac{1}{6}}+...\\
        \Gamma^{0,1}_{4,4}\chvec{0}{0}
        =&q^{-\frac{1}{6}}\bar q^{-\frac{1}{6}}+...,       &&\Gamma^{0,1}_{4,4}\chvec{0}{1}=q^{-\frac{1}{6}}\bar q^{-\frac{1}{6}}+....
\end{aligned}
\end{equation}

\subsubsection{\texorpdfstring{$h^2$}{h2}-twisted RR sector}
We take $a=1$ (then $b=0$ since $\vartheta_{11}=0$) in equation \eqref{eq:h^2_twisted_trace}:

\begin{equation}
    \begin{aligned}
        &\frac{1}{16(\Im\tau)^2\eta^8\bar \eta^8}
\sum_{c,d,l=0}^1 
\left|\tchar{1}{0}^2\tchar{1+c}{d+l}
\tchar{1-c}{-d}\right|^2
\Gamma_{4,4}^{0,l}\chvec{c}{d}\\
        =&\frac{1}{16(\Im \tau)^2\eta^{8}\bar\eta^{8}}\left|\vartheta\chvec{1}{0}^4\right|^2\Gamma^{0,0}\chvec{0}{0}\quad (c=0 \text{ then } (d,l)=(0,0))\\
        &\quad+\frac{1}{16(\Im \tau)^2\eta^{8}\bar\eta^{8}}\sum_{d,l=0}^1 \left|\vartheta\chvec{1}{0}^2\vartheta\chvec{0}{d+l}\vartheta\chvec{0}{d}\right|^2\Gamma^{0,0}\chvec{1}{d}\quad (c=1 \text{ then } l=0)\\
         =&\frac{1}{16(\Im \tau)^2\eta^{8}\bar\eta^{8}}\qty(|\vartheta_{10}^4|^2\Gamma^{0,0}\chvec{0}{0}
        +\sum_{d=0}^1 \left|\vartheta_{10}^2\vartheta_{0d}^2\right|^2\frac{|\eta|^4}{\left|\vartheta\chvec{0}{1+d}\right|^2})\\
       =&\frac{1}{(\Im \tau)^2}\qty(48+...).
    \end{aligned}
\end{equation}
These  $48$ RR massless states can be seen as $U(1)^{12}$ vectors in IIA, or $4$ scalars and $12$ SD/ASD tensors in IIB in the light-cone formalism. These additional RR fields are similar to the additional RR fields in Type 0A/0B theories.

\subsection{Spectrum and moduli space}

The massless and tachyonic spectra are summarized in Tables \ref{table:IIAspectrum_Enr} and \ref{table:IIBspectrum_Enr}. Unlike supersymmetric cases, it is known that there are several subtle points in non-supersymmetric spectrum \cite{Bergman:1999km}. 

\begin{table}[h]
\centering
\renewcommand{\arraystretch}{1.4}
\setlength{\dashlinedash}{1.2pt}
\setlength{\dashlinegap}{1.2pt}
\setlength{\arrayrulewidth}{0.8pt}

\begin{tabular}{c|c:c|c|c:c|c|}
 & \multicolumn{2}{c|}{untwisted} & $h$-twisted & \multicolumn{2}{c|}{$h^{2}$-twisted} & $h^{3}$-twisted \\
\hline
NSNS
&\begin{tabular}[c]{@{}l@{}}
$g_{\mu\nu},B_{\mu\nu},\phi$  \\
$8$ scalars 
\end{tabular}
& $32$ scalars
&$4$ tachyons & one tachyon & 32 scalars& $4$ tachyons \\
\hline
RR
& \begin{tabular}[c]{@{}l@{}}
$C_{\mu},C_{\mu 67}$\\
$C_{\mu 89},C_{\mu\nu\rho}$
\end{tabular}
& $8$ vectors
& massive & $4$ vectors &$8$ vectors & massive \\
\hline
\end{tabular}\caption{Massless and tachyonic spectrum of IIA on $\text{Enriques}=K3/\Z_2$. The dotted line separates the origin (untwisted or $g$-twisted sector of  IIA on $K3=T^4/\Z_2$). Except for the $32$ scalars in $h^2$-twisted sector and one half of RR fields, there are 41 massless scalars and $12$ vectors ($U(1)^{12}$). The tachyon in the $h^2$-twisted sector is moduli-independent, but it would also become massive at strong coupling \cite{Bergman:1999km}. Tachyons in $h,h^3$-twisted sectors are moduli-dependent.}\label{table:IIAspectrum_Enr}
\end{table}

\begin{table}[h]

\centering
\renewcommand{\arraystretch}{1.4}
\setlength{\dashlinedash}{1.2pt}
\setlength{\dashlinegap}{1.2pt}
\setlength{\arrayrulewidth}{0.8pt}

\begin{tabular}{c|c:c|c|c:c|c|}
 & \multicolumn{2}{c|}{untwisted} & $h$-twisted &  \multicolumn{2}{c|}{$h^{2}$-twisted} & $h^{3}$-twisted \\
\hline
NSNS
&\begin{tabular}[c]{@{}l@{}}
$g_{\mu\nu},B_{\mu\nu},\phi$  \\
$8$ scalars 
\end{tabular}
& $32$ scalars
& $4$ tachyons  & a tachyon & 32 scalars& $4$ tachyons \\
\hline
RR
& \begin{tabular}[c]{@{}l@{}}
$C,C_{67},C_{89},C^+_{6789}$\\
$C_{\mu \nu},C^+_{\mu\nu 67}, C^-_{\mu\nu 89}$
\end{tabular}
& \begin{tabular}[c]{@{}l@{}}
$8$ scalars\\
$8 $ ASD tensors
\end{tabular}
& massive &  \begin{tabular}[c]{@{}l@{}}
$4$ scalars \\
 $2$ ASD / $2 $ SD tensors
\end{tabular}& \begin{tabular}[c]{@{}l@{}}
$8$ scalars \\
   $8 $ SD tensors
\end{tabular} & massive \\
\hline
\end{tabular}\caption{Massless and tachyonic spectrum of IIB on $\text{Enriques} =K3/\Z_2$. The dotted line separates the origin (untwisted or $g$-twisted sector of  IIB on $K3=T^4/\Z_2$).  Except for $32$ scalars in $h^2$-twisted sector and one half of RR fields, there are $53$ scalar fields and $7$ ASD and $7$ SD tensors (then $U(1)^{14}$ gauge symmetry). Because the numbers of SD/ASD tensors match, the gravitational anomaly cancels. The tachyon in the $h^2$-twisted sector is moduli-independent, while tachyons in $h,h^3$-twisted sectors are moduli-dependent.}\label{table:IIBspectrum_Enr}
\end{table}

Compared to a K3 surface, whose homology lattice $H_0\oplus H_2\oplus H_4$ is $(4,20)$, an Enriques surface has half of that, namely the $(2,10)$ lattice. Therefore, when one considers its decomposition, it is natural for the moduli space to take the following form:

\begin{equation}\label{eq:moduli_of_II_Enriques}
\begin{aligned}
    \mathcal{M}^{\text{IIA}}_{\text{Enriques}}=&\frac{O(2,10)}{O(2)\times O(10)}\times \frac{O(2,10)}{O(2)\times O(10)}\times \R_{\text{dilaton}},\\
    \mathcal{M}^{\text{IIB}}_{\text{Enriques}}=&\frac{O(3,11)}{O(3)\times O(11)}\times \frac{O(2,10)}{O(2)\times O(10)}.
\end{aligned}
\end{equation}
Dilaton is included in the latter. Later in subsection \ref{subsec:type0_gravity}, we will see that the same result follows also from the moduli space of Ricci-flat metrics on Enriques surfaces, in terms of target gravity.

The dimensions of the moduli spaces are equal to the numbers of massless scalar fields in the theories:
\begin{equation}
    \begin{aligned}
        \mathcal{M}^{\text{IIA}}_{\text{Enriques}} =&41,\\
        \mathcal{M}^{\text{IIB}}_{\text{Enriques}}=&53.
    \end{aligned}
\end{equation}

\subsection{On the spin structure of Enriques surfaces}
According to Rokhlin's theorem, an Enriques surface does not admit a spin structure, since its $H_2$ lattice has signature $(1,9)$. One might therefore worry that compactifying Type II theories (which contain spacetime fermions) on an Enriques surface is inconsistent from the viewpoint of the target-space supergravity, in spite of the consistency of the worldsheet construction \cite{Acharya:2019mcu}. In fact, because there are no perturbative fermions in the spectrum, the decompactification limit does not yield Type II supergravity: instead, it is described by Type 0 gravities \cite{Dixon:1986iz,SEIBERG1986272,Meessen_2001,Klebanov_1999, Sharpe:2013bwa}, which are the low-energy limit of Type 0 strings.  In the next section, we will explain that the theories constructed in this section can be viewed as Type 0 strings compactified on an Enriques surface.

\newpage
\section{Type 0 superstring theories compactified on a K3/Enriques surface}\label{sec:Type 0}
In this section, we consider another aspect of the non-supersymmetric theories obtained in the previous section. We will find that low-energy limits of these two theories would be described by Type 0 gravitational theories. The lift of Type 0A to M-theory would provide a non-perturbative argument on the spectra.

\subsection{Type 0 theories in ten dimensions}

It is well known that there are non-supersymmetric cousins of Type II superstrings, called Type 0 strings \cite{Dixon:1986iz,SEIBERG1986272,Bergman:1999km}. Type 0 superstring theories can be constructed in two different ways: (1) diagonal GSO projection, or (2) $(-1)^F$ orbifolding from Type II superstring theories, where $F$ is the spacetime fermion number.

After the GSO projection, the Hilbert spaces for Type II and Type 0 theories are given as follows:
\begin{equation}
    \begin{aligned}
        \text{IIA} =&\text{(NS+, NS+)$\oplus$(NS+, R+)$\oplus$(R+, NS+)$\oplus$(R+,R-)},\\
        \text{IIB} =&\text{(NS+, NS+)$\oplus$(NS+, R+)$\oplus$(R+, NS+)$\oplus$(R+,R+)},\\
        \text{0A}=&\text{(NS+, NS+)$\oplus$(NS-, NS-)$\oplus$(R+, R-)$\oplus$(R-,R+)},\\
        \text{0B}=&\text{(NS+, NS+)$\oplus$(NS-, NS-)$\oplus$(R+,R+)$\oplus$(R-,R-)}.
    \end{aligned}
\end{equation}
It can be seen that there is no $(\text{NS}_+,\text{R}_\pm)$ or $(\text{R}_{\pm},\text{NS}_+)$ sector in Type 0 theories, which means there are no spacetime fermions perturbatively. There is a tachyon in $(\text{NS}_-,\text{NS}_-)$. 

Just as Type IIA theory is obtained by compactifying M-theory on a circle, it is also conjectured that Type 0A theory can be obtained in the following way \cite{Bergman:1999km}:
\begin{equation}
    \text{Type 0A}= \text{M-theory on } S^1/S\cdot (-1)^F,
\end{equation}
where $S$ is a half-shift along the $S^1$, and $F$ is the spacetime fermion number. Assuming this relation, it is also conjectured in \cite{Bergman:1999km} that (1) the tachyon in Type 0A theory would become massive at strong coupling, and (2) one half of the RR fields would also become massive at non-zero coupling constant.

\subsection{K3 surfaces and Enriques surfaces}
It is useful to view Type 0 theories as the $(-1)^F$ orbifold of Type II theories. Let $\H_{T^4}$ be the Hilbert space of Type II theories on $T^4$. The Hilbert space of Type 0 theories on $T^4$ can be written as 

\begin{equation}
\begin{aligned}
    \H^{\text{Type 0}}_{T^4}=&\frac{1+(-1)^F}{2}\H_{T^4} &&\text{(untwisted sector)}\\
    \oplus&\frac{1+(-1)^F}{2} \H^{(-1)^F}_{T^4},  &&\qty(\text{$(-1)^F$-twisted sector})
\end{aligned}
\end{equation}
where $\H^{(-1)^F}_{T^4}$ is the $(-1)^F$-twisted Hilbert space. Next, we can obtain new Hilbert space by gauging $g$:
\begin{equation}
    \begin{aligned}
        \H^{\text{Type 0}}_{K3}=&\frac{1+g}{2}\frac{1+(-1)^F}{2}\qty(\H_{T^4}\oplus \H^{(-1)^F}_{T^4}),\quad \qty(\text{untwisted sector})\\
        \oplus&\frac{1+g}{2}\frac{1+(-1)^F}{2}\qty(\H^g_{T^4}\oplus \H^{g(-1)^F}_{T^4})\quad \qty(\text{$g$-twisted sector})
    \end{aligned}
\end{equation}
This is the Hilbert space of Type 0 theories on $K3=T^4/g$ .
Finally, by gauging $h$, we obtain the Hilbert space of Type 0 on an Enriques surface ($(T^4/g)/h$) as follows:
\begin{equation}
    \begin{aligned}
        \H^{\text{Type 0}}_{\text{Enriques}}=&\frac{1+h}{2}\frac{1+g}{2}\frac{1+(-1)^F}{2}\qty(\H_{T^4}\oplus \H^{(-1)^F}_{T^4}\oplus\H^g_{T^4}\oplus \H^{g(-1)^F}_{T^4})\\
        \oplus&\frac{1+h}{2}\frac{1+g}{2}\frac{1+(-1)^F}{2}\qty(\H^h_{T^4}\oplus \H^{h(-1)^F}_{T^4}\oplus\H^{gh}_{T^4}\oplus \H^{gh(-1)^F}_{T^4}).
    \end{aligned}
\end{equation}
The first line is the untwisted sector, while the second line is the $h$-twisted sector. Since there are no fermionic states in Type 0 theories, $h$ acts as a $\Z_2$ operator on Type 0 Hilbert spaces. Note that this is the same Hilbert space of Type II theories on an Enriques surface (equation \eqref{eq:HilEnr}), with $h^2=(-1)^F$.  Thus, we conclude that Type 0 superstring theories on $ K3/h$ can be seen as $h$ -orbifold Type II superstring theories on $K3$: 

\begin{equation}
    \text{Type 0 on } \frac{K3}{h}=\frac{\text{Type II on K3}}{h}.
\end{equation}

\subsection{Type 0 gravitational theories}\label{subsec:type0_gravity}
In the previous section, we identified the moduli spaces of Type II on $K3/h$. In this subsection, we identify the moduli spaces of the theories in terms of low-energy Type 0 gravity.

It is known that the moduli space of Ricci-flat metrics on an Enriques surface is \cite{Itoh1994}:
\begin{equation}
    \M^{\text{flat}}_{\text{Enriques}}=\R_{\text{Volume}}\times \frac{O(1,9)}{O(1)\times O(9)}\times\frac{O(2,10)}{O(2)\times O(10)},
\end{equation}
which is $30$-dimensional. Then, the classical moduli spaces can be described in terms of $ \M^{\text{flat}}_{\text{Enriques}}$:
\begin{equation}
\begin{aligned}
\mathcal{M}^{\text{0A Gravity}}_{\text{Enriques}}
=&\M^{\text{flat}}_{\text{Enriques}} \times\R^{1,9}_{B}\times\R_{\text{dilaton}},\\
\mathcal{M}^{\text{0B Gravity}}_{\text{Enriques}}
=&\M^{\text{flat}}_{\text{Enriques}}\times \R^{1,9}_B\times \qty(\R^{2,10}_{RR})^2\times \R_{\text{dilaton}},
\end{aligned}
\end{equation}

\begin{comment}
    by using the following equations:
\begin{equation}
    \begin{aligned}
        \
        \frac{O(3,11)}{O(3)\times O(11)}=&\R_{\text{Volume}}\times \frac{O(2,10)}{O(2)\times O(10)}\times \R^{2,10}_{RR}\\
=&\R_{\text{Volume}}\times \frac{O(1,9)}{O(1)\times O(9)}\times\R_{\text{dialton}}\times \R^{1,9}_B\times \R^{2,10}_{RR}.
    \end{aligned}
\end{equation}
\end{comment}
$\R^{1,9}_B$ corresponds to the $H_2$ lattice of an Enriques surface, which would be wrapped by B-fields, while $\R^{2,10}_{RR}$ corresponds to the homology lattice $H_0\oplus H_2\oplus H_4$, which would be wrapped by even-form RR fields. Note that the RR fields in Type 0 gravity are doubled.

It can be checked that $\dim\M^{\text{0A Gravity}}_{\text{Enriques}}$ matches the $\mathcal{M}^{\text{IIA}}_{\text{Enriques}}$ we saw in equation \eqref{eq:moduli_of_II_Enriques} by using the following relation:
\begin{equation}
    \R_{\text{Volume}}\times \frac{O(1,9)}{O(1)\times O(9)}\times \R^{1,9}_{B}=\frac{O(2,10)}{O(2)\times O(10)}.
\end{equation}
On the other hand, $\dim\M^{\text{0B Gravity}}_{\text{Enriques}}$ differs from $\mathcal{M}^{\text{IIB}}_{\text{Enriques}}$ by $\R^{2,10}_{RR}$. It would be explained by the non-perturbative lifting of the masses of half of the RR fields we discussed earlier.

\newpage
\section{Non-supersymmetric dualities from supersymmetric ones}\label{sec:dualities}

\subsection{Via K3 surfaces (supersymmetric)}
Before proceeding, we review the supersymmetric string-string dualities in six and five dimensions. 
\subsubsection{In six dimensions}

It is known that  the heterotic string on $T^4$ is dual to the Type IIA compactified on a K3 surface \cite{Witten_1995,Aspinwall:1994rg}. The moduli spaces are the same:

\begin{equation}
\begin{aligned}
    \mathcal{M}^{\text{IIA}}_{\text{K3}}=&\frac{O(4,20)}{O(4)\times O(20)}\times\R_{\text{dilaton}},\\
\end{aligned}
\end{equation}

\begin{equation}
\begin{aligned}
    \mathcal{M}^{\text{Het}}_{T^4}=&\frac{O(4,20)}{O(4)\times O(20)}\times\R_{\text{dilaton}},\\
\end{aligned}
\end{equation}
which means that both theories have $\dim \M=81$ massless scalar fields. $64$ Massless scalar states in the $g$-twisted sector in Type II side correspond to the $64$ Wilson lines in the heterotic side. There are many other common points: supersymmetry, low-energy supergravity, and enhanced gauge symmetry. Some properties can be understood by non-perturbative argument in terms of BPS states.

\begin{comment}

Corresponding values of Wilsonlines on heterotic side are given by \cite{Kiritsis_2000}

\begin{equation}
    y=\frac{1}{2}
\begin{pmatrix}
0101 & 0101 & 0101 & 0101\\
0000 & 0000 & 1111 & 1111\\
0000 & 1111 & 0000 & 1111\\
0011 & 0011 & 0011 & 0011
\end{pmatrix}=
\begin{pmatrix}
A_1 \\ A_2 \\ A_3\\ A_4 
\end{pmatrix}
\end{equation}

\begin{equation}
     A_i\cdot A_j=\begin{cases}
         2,i=j\\
         1,i\neq j
     \end{cases}
\end{equation}

\begin{equation}
l_\text{H}=g_{\text{6IIA}}l_{\text{II}},\quad g_{\text{6IIA}}=\frac{1}{g_{\text{6H}}},\quad \qty(\frac{R_H}{l_\text{H}})^2=\frac{V_{K3}}{l^4_{\text{II}}}
\end{equation}

\begin{equation}
    \begin{aligned}
        V_{K3}=R_{\text{H}6}^2,\quad G_{3\text{IIA}}=G_{3\text{H}},\quad \qty(\frac{G^{96}}{G^{99}},\frac{G^{97}}{G^{99}},\frac{G^{98}}{G^{99}})_{3\text{IIA}}=(B_{67},B_{78},B_{89})_{\text{H}}\\
        B_{16}=A_6,
    \end{aligned}
\end{equation}
\end{comment}

\subsubsection{In five dimensions}

It is also known that the strong coupling limit of the heterotic string on $T^5$ is given by Type IIB compactified on a K3 surface \cite{Witten_1995}. The moduli spaces support the conjecture:
\begin{equation}
\begin{aligned}
    \mathcal{M}^{\text{IIB}}_{\text{K3}}=&\frac{O(5,21)}{O(5)\times O(21)},\\
    \mathcal{M}^{\text{Het}}_{T^5}=&\frac{O(5,21)}{O(5)\times O(21)}\times \R_{\text{dilaton}}.
\end{aligned}
\end{equation}
Both theories have $16$ supercharges and rank $26$ gauge symmetries.

\subsection{Via Enriques surfaces (non-supersymmetric)}

In section \ref{sec:Enr}, we considered the orbifold of Type II and Type 0 theories by the involution $h$ on a $K3$ surface. It is known that this involution induces an action on a lattice $H_2(K3;\Z)=2E_8\oplus 3\Gamma_{1,1}$ as follows:

\begin{equation}
\begin{aligned}
        h:&\qty(x_1^{(E_8)},x_2^{(E_8)},y_1,y_2,y_3)\\
        \mapsto&\qty(x_2^{(E_8)},x_1^{(E_8)},y_2,y_1,-y_3).
\end{aligned}
\end{equation}
This is the same action we saw in the heterotic situation in section \ref{sec:het}.

\begin{figure}[h]
    \centering

    \begin{tikzcd}[row sep=huge, column sep=huge]
\text{Type IIA on }K3 
  \arrow[rr, leftrightarrow, "\mathrm{dual}"] 
  \arrow[d, "h"']
&&
\text{Het on }T^4 
  \arrow[d, "h"]
\\
\text{Type 0A on Enriques} 
  \arrow[rr, leftrightarrow, "\mathrm{dual}"] 
&&
\text{Het on }T^4/\mathbb{Z}_2
\end{tikzcd}

    \begin{tikzcd}[row sep=huge, column sep=huge]
\text{Type IIB on }K3 
  \arrow[rr, leftarrow, "\mathrm{strong~ coupling ~limit}"] 
  \arrow[d, "h"']
&&
\text{Het on }T^5
  \arrow[d, "h"]
\\
\text{Type 0B on Enriques} 
  \arrow[rr, leftarrow, "\mathrm{strong~ coupling ~limit}"] 
&&
\text{Het on }T^5/\mathbb{Z}_2
\end{tikzcd}
    \caption{A schematic picture.}\label{fig:schematic2}
\end{figure}
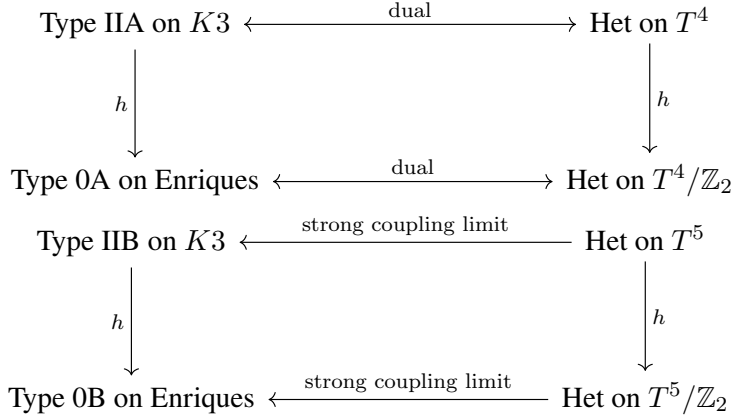

Assuming the previous supersymmetric dualities with $K3$ surfaces, taking its quotient by $h$ suggests two new non-supersymmetric dualities: the heterotic string on $T^4/h$ is dual to Type 0A on an Enriques surface, and the strong coupling limit of the heterotic string on $T^5/h$ is given by Type 0B on an Enriques surface \ref{fig:schematic2}.

The main underlying assumption is that the duality map commutes with the quotient by $h$. In practice, this means that the $h$-action on the Narain lattice on the heterotic side matches the $h$-action induced by the fixed-point-free involution on the K3 geometry (and hence on the CFT) on the Type II side. Under this assumption, the resulting orbifold theories should again share the same moduli spaces:
\begin{equation}
\begin{aligned}
\mathcal{M}^{\text{IIA}}_{\text{Enriques}}=&\frac{O(2,10)}{O(2)\times O(10)}\times \frac{O(2,10)}{O(2)\times O(10)}\times \R_{\text{dilaton}},\\
   \M^{\text{Het}}_{T^4/h}= &\frac{O(2,10)}{O(2)\times O(10)}\times\frac{O(2,10)}{O(2)\times O(10)}\times \R_{\text{dilaton}},
   \end{aligned}
\end{equation}
and 

\begin{equation}
\begin{aligned}
        \mathcal{M}^{\text{IIB}}_{\text{Enriques}}=&\frac{O(3,11)}{O(3)\times O(11)}\times \frac{O(2,10)}{O(2)\times O(10)},\\
   \M^{\text{Het}}_{T^5/h}=&\frac{O(3,11)}{O(3)\times O(11)}\times\frac{O(2,10)}{O(2)\times O(10)}\times \R_{\text{dilaton}}.
\end{aligned}
\end{equation}
Note that the factor $\R_{\text{dilaton}}$ in $\M^{\text{Het}}_{T^5/h}$ would drop in the strong coupling limit.

We can compare the massless spectra. First of all, there are no uncharged fermions (dilatino and gravitino) on both theories. Another point is that the rank of gauge symmetries, which is $12$ for 0A and $14$ for 0B, also matches that of heterotic strings. It should be assumed that $32$ NSNS massless states in the $h^2$-twisted sector on the Type II side, which correspond to the higgsed $32$ Wilson lines on the heterotic side, would become massive at strong coupling.

\subsection{On tachyons}

The tachyon structure differs between the Type 0 and heterotic descriptions. First, a moduli-independent tachyon appears on the Type 0 side, whereas no such tachyon is present perturbatively on the heterotic side. Second, the regions of moduli space where moduli-dependent tachyons appear do not coincide.

We interpret these discrepancies as follows. Assuming the M-theory lift of Type 0A, the moduli-independent Type 0 tachyon is expected to become massive at strong coupling \cite{Bergman:1999km}. Since weakly coupled heterotic strings correspond to strongly coupled Type 0 theories in the proposed duality, this suggests a non-perturbative instability on the heterotic side throughout moduli space. Moreover, the Type 0 tachyon mass can be shifted by RR fluxes in the target-space effective action \cite{Klebanov_1999}, which means that the worldsheet computation is not enough to investigate tachyons.

\acknowledgments
The author thanks Jin Miyazawa, Yuta Hamada, Kantaro Ohmori, Masashi Kawahira, Justin Kaidi, Hiroki Wada, Yuichi Koga, Taro Kimura for useful comments and discussions.

\begin{appendix}

\section{Theta functions and \texorpdfstring{$D_4$}{D4} characters}\label{sec:theta}
In this appendix, we summarize the formulas used in the paper.
Let $\tau$ be a complex number with positive imaginary part, and $q=\exp \left(2\pi i \tau\right)$. 

\subsection{Theta functions}
The theta functions are

\begin{equation}\label{eq:thetas}
\begin{aligned}
\vartheta_1=\vartheta\Bigl[\begin{matrix}1\\1\end{matrix}\Bigr]\coloneqq &i \sum_{n \in \mathbb{Z}}(-1)^n q^{\frac{1}{2}\qty(n-\frac{1}{2})^2}=0, \\
\vartheta_2=\vartheta\Bigl[\begin{matrix}1\\0\end{matrix}\Bigr]\coloneqq&\sum_{n \in \mathbb{Z}} q^{\frac{1}{2}\qty(n-\frac{1}{2})^2}  =2 q^{\frac{1}{8}} \prod_{n=1}^{\infty}\qty(1-q^n)\qty(1+ q^n)\qty(1+ q^n),\\
\vartheta_3=\vartheta\Bigl[\begin{matrix}0\\0\end{matrix}\Bigr]\coloneqq&\sum_{n \in \mathbb{Z}} q^{\frac{1}{2} n^2} =\prod_{n=1}^{\infty}\qty(1-q^n)\qty(1+ q^{n-\frac{1}{2}})\qty(1+ q^{n-\frac{1}{2}}),\\
\vartheta_4=\vartheta\Bigl[\begin{matrix}0\\1\end{matrix}\Bigr]\coloneqq&\sum_{n \in \mathbb{Z}}(-1)^n q^{\frac{1}{2} n^2}=\prod_{n=1}^{\infty}\qty(1-q^n)\qty(1- q^{n-\frac{1}{2}})\qty(1-q^{n-\frac{1}{2}}).
\end{aligned}
\end{equation}
A more general definition is given by

\begin{equation}
    \tchar{a}{b}=\sum_{n\in\Z}q^{\frac{1}{2}\qty(n+\frac{a}{2})^2}e^{\pi i b\qty(n+\frac{a}{2})}, a,b\in\Z.
\end{equation}

\subsection{Properties of theta functions}

We used the following equations elsewhere:

\begin{equation}\label{eq:theta_priodicity}
\begin{aligned}
\vartheta\chvec{a\pm 2}{b}&=\vartheta\chvec{a}{b},&&\vartheta\chvec{a}{b\pm 2}=e^{\pm\pi i a}\vartheta\chvec{a}{b},\\
T\cdot\vartheta\chvec{a}{b}=&e^{-\frac{1}{4}\pi i a(a-2)}\vartheta\chvec{a}{b},&&S\cdot\vartheta\chvec{a}{b}=\qty(-i\tau)^{\frac{1}{2}}e^{\frac{1}{2}\pi i ab}\vartheta\chvec{b}{-a},\\
\end{aligned}
\end{equation}

\begin{equation}
    \prod_{n=1}^\infty\frac{1}{(1+q^n)^4}=\frac{4\eta^6}{\vartheta_{10}^2}\prod_{n=1}\frac{1}{(1-q^n)^4}.
\end{equation}

\subsection{\texorpdfstring{$D_4$}{D4} characters}\label{subsec:characters}
The vector and spinor conjugacy classes of $D_4$ are
\begin{equation}
\begin{aligned}
        V=&\left\{(n_1, \cdots, n_4) \mid n_i \in \mathbb{Z}, \sum_{i=1}^4 n_i \in 2 \mathbb{Z}+1\right\} ,\\
        Sp=&\left\{\left(n_1+\frac{1}{2}, \cdots, n_4+\frac{1}{2}\right) \right\rvert n_i \in \mathbb{Z}, \sum_{i=1}^4 n_i \in 2 \mathbb{Z}\} ,
\end{aligned}
\end{equation}
and their lattice theta functions are 
\begin{equation}
    \begin{aligned}
        \sum_{p\in V}q^{\frac{1}{2}p^2}=&\theta_{00}^4-\theta_{01}^4,\\
        \sum_{p\in Sp}q^{\frac{1}{2}p^2}=&\theta_{10}^4+\theta_{11}^4.
    \end{aligned}
\end{equation}

\section{Twisted bosonic Fock space}\label{sec:Twisted bosonic fock space}

In this section we construct $g,h,gh$-twisted Hilbert spaces for $\H_{\Gamma_{4,4}}$ explicitly, and calculate $\Gamma^{k,l}\chvec{c}{d}=\tr_{\H_{\Gamma_{4,4}}^{g^c h^k}}g^d h^l\q$ as its trace. On the compactification of Type II strings on $T^4$, there is a Hilbert space associated with $\Gamma_{4,4}=H^1(T^4;\Z)\oplus H_1(T^4;\Z)$:
\begin{equation}
\H_{\Gamma_{4,4}}=\H_{\text{Boson}}^{4,4}\otimes\qty(\bigoplus_{p\in\Gamma_{4,4}}\C \ket{p}).
\end{equation}
The operators $g,h$ act $\ket{p}$ as follows:
\begin{equation}
    \begin{aligned}
        g\ket{p_6,p_7,p_8,p_9}=&\ket{-(p_6,p_7,p_8,p_9)},\\
        h\ket{p_6,p_7,p_8,p_9}=&(-1)^{R_8 p_8}\ket{-p_6,-p_7,p_8,p_9},\\
gh\ket{p_6,p_7,p_8,p_9}=&(-1)^{R_7 p_7}\ket{p_6,p_7,-p_8,-p_9}.\\
    \end{aligned}
\end{equation}
Now we can compute $\Gamma^{0,l}\chvec{0}{d}$ as follows:

\begin{equation}
    \begin{aligned}
     \Gamma_{4,4}^{0,0}\chvec{0}{0}=&\tr_{\H_{\Gamma_{4,4}}}q^{L_0}\bar{q}^{\bar L_0}
     =\frac{1}{\eta^4\bar \eta^4}\sum_{p\in\Gamma_{4,4}}q^{\frac{1}{2}p_L^2}\bar q^{\frac{1}{2}p_R^2},\\
     \Gamma_{4,4}^{0,0}\chvec{0}{1}=&\tr_{\H_{\Gamma_{4,4}}}g q^{L_0}\bar{q}^{\bar L_0}
     =\frac{16|\eta|^4}{|\vartheta_{10}|^4},\\
    \Gamma^{0,1}_{4,4}\chvec{0}{0}=&\tr_{\H_{\Gamma_{4,4}}}hq^{L_0}\bar{q}^{\bar L_0}\\
        =&\frac{1}{\eta^4\bar\eta^4}\biggl|\frac{2\eta^3}{\vartheta_{10}}\biggr|^2\sum_{n\in\Z}\sum_{w\in \Z}(-1)^nq^{\frac{1}{2}\qty(\frac{n}{R_8}+wR_8)^2}\bar q^{\frac{1}{2}\qty(\frac{n}{R_8}-wR_8)^2}\vartheta_{\Gamma_{1,1}(R_9)},\\
        \Gamma^{0,1}_{4,4}\chvec{0}{1}=&\tr_{\H_{\Gamma_{4,4}}}gh q^{L_0}\bar{q}^{\bar L_0}\\
        =&\frac{1}{\eta^4\bar\eta^4}\biggl|\frac{2\eta^3}{\vartheta_{10}}\biggr|^2\vartheta_{\Gamma_{1,1}(R_6)}\sum_{n\in\Z}\sum_{w\in \Z}(-1)^nq^{\frac{1}{2}\qty(\frac{n}{R_7}+wR_7)^2}\bar q^{\frac{1}{2}\qty(\frac{n}{R_7}-wR_7)^2},
    \end{aligned}
\end{equation}
where $\Gamma_{1,1}(R)$ is a $(1,1)$ even self-dual lattice and  $\vartheta_{\Gamma_{1,1}(R)}$ is its theta function:

\begin{equation}
\begin{aligned}
\Gamma_{1,1}(R)=&\frac{1}{\sqrt{2}}\{(\frac{n}{R}+wR;\frac{n}{R}-wR)|n,w\in\Z\},\\
\vartheta_{\Gamma_{1,1}(R)}=&\sum_{p\in\Gamma_{1,1}(R)}q^{\frac{1}{2}p_L^2}q^{\frac{1}{2}p_R^2}.
\end{aligned}
\end{equation}

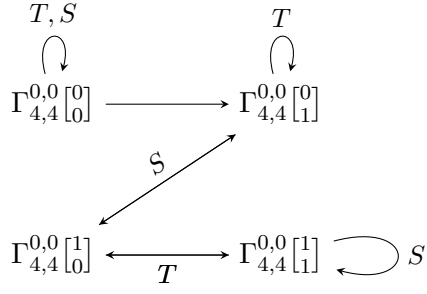
\begin{figure}
    \centering
    \begin{tikzpicture}[>=stealth, baseline=(current bounding box.center)]

% Nodes
\node (A) at (0,2) {$\Gamma_{4,4}^{0,0}\chvec{0}{0}$};
\node (B) at (3,2) {$\Gamma_{4,4}^{0,0}\chvec{0}{1}$};
\node (C) at (0,0) {$\Gamma_{4,4}^{0,0}\chvec{1}{0}$};
\node (D) at (3,0) {$\Gamma_{4,4}^{0,0}\chvec{1}{1}$};

% Arrows
\draw[->] (A) edge[loop above] node[above] {\small $T, S$} (A);
\draw[->] (B) edge[loop above] node[above] {\small $T$} (B);
\draw[->] (D) edge[loop right] node[right] {\small $S$} (D);

\draw[->] (A) -- (B) node[midway, above] {};
\draw[->] (C) -- (D) node[midway, below] {\small $T$};
\draw[->] (D) -- (C) node[midway, below] {\small $T$};

\draw[->] (B) -- (C) node[midway, sloped, above] {\small $S$};
\draw[->] (C) -- (B) node[midway, sloped, below]{} ;

\end{tikzpicture}
    \caption{Modular transformations of partition function on $T^4/\Z_2$}
\end{figure}

\subsection{\texorpdfstring{$g$}{g}-twisted space}
The boundary conditions of $g$-twisted bosons $X^i_g:T^2\to T^4$ are 

  \begin{equation}
    \begin{aligned}
    X_g^{i}(\sigma_1+2\pi,\sigma_2)=&g\cdot X_g^i(\sigma_1,\sigma_2)\\
        =&-X_g^i(\sigma_1,\sigma_2),\quad6\leq i,j\leq 9.
    \end{aligned}
    \end{equation}

Let us denote the Fock space of $X_g^i$ as $\H^{4,4}_{X_g}$. The $g$-twisted Hilbert space $\H_{\Gamma_{4,4}}^{g}$ is given concretely as follows: 
\begin{equation}
\begin{aligned}
\H_{\Gamma_{4,4}}^{g}=&\bigoplus_{\text{fixed points}}\qty(\H^{4,4}_{X_g}\otimes\C\ket{x_6,x_7,x_8,x_9})\\
=&16\H^{4,4}_{X_g},
\end{aligned}
\end{equation}
where all 16 states $\ket{x_6,x_7,x_8,x_9}$ are $g$ invariant, while one half of them are $h$ invariant.

    Then it holds that

    \begin{equation}
\tr_{\H_{\Gamma_{4,4}}^{g}}\q=S^{\pm1}\cdot \tr_{\H_{\Gamma_{4,4}}}g \q
    \end{equation}

    \begin{equation}
        \begin{aligned}
            [\alpha^i_r,\alpha^j_s]=&[\tilde{\alpha}^i_r,\tilde{\alpha}^j_s]=r\delta^{ij}\delta_{r+s,0},\\
            [\alpha^i_r,\tilde{\alpha}^j_s]=&0, \quad r,s\in\Z+\frac12,\quad6\leq i,j\leq 9.
        \end{aligned}
    \end{equation}

    \begin{equation}
        \begin{aligned}
            L_0=&\frac{1}{2}\sum_{i=6}^9\sum_{r=1/2}\qty(\alpha_{-r}^i\alpha_{r}^i+\alpha_r^i\alpha_{-r}^i)\\
            =&\sum_{i=6}^9\sum_{r=1/2}\alpha_{-r}^i\alpha_{r}^i+\frac{1}{12}.
        \end{aligned}
    \end{equation}

    \begin{equation}\label{eq:gamma01}
        \begin{aligned}
            \Gamma_{4,4}^{0,0}\chvec{1}{0}=&\tr_{\H^{g}_{\Gamma_{4,4}}} \q
            =16\biggl|\frac{\eta^4}{\vartheta_{01}^4}\biggr|,\\
        \Gamma_{4,4}^{0,0}\chvec{1}{1}=&\tr_{\H^{g}_{\Gamma_{4,4}}}g \q
        =16\biggl|\frac{\eta^4}{\vartheta_{00}^4}\biggr|,\\
    \Gamma_{4,4}^{0,1}\chvec{1}{0}=&\tr_{\H^{g}_{\Gamma_{4,4}}} h \q
        =0,\\
    \Gamma_{4,4}^{0,1}\chvec{1}{1}=&\tr_{\H^{g}_{\Gamma_{4,4}}} gh \q
        =0.
    \end{aligned}
\end{equation}

\subsection{\texorpdfstring{$h$}{h}-twisted space}

The boundary conditions of $h$-twisted bosons $X^i_h:T^2\to T^4$ are given as follows:
 \begin{equation}
    \begin{aligned}
    X_h^{i}(\sigma_1+2\pi,\sigma_2)=&h\cdot X_h^i(\sigma_1,\sigma_2),i=6,7,8,9,
    \end{aligned}
    \end{equation}
where
 \begin{equation}
             h :
\begin{cases}
X^6 \rightarrow -X^6,\\
X^7 \rightarrow -X^7 + \pi R_7 ,\\
X^8 \rightarrow X^8 + \pi R_8,\\
X^9 \rightarrow X^9. \\
\end{cases}
 \end{equation}

Let us denote the Fock space of $X_h^i$ as $\H^{4,4}_{X_h}$.  The $h$-twisted Hilbert space $\H_{\Gamma_{4,4}}^{h}$ is given concretely as follows: 

\begin{equation}
\begin{aligned}
\H_{\Gamma_{4,4}}^{h}=&
\H^{4,4}_{X_h}\otimes \left(\bigoplus_{\substack{x^6=0,\pi R_6 \\ x^7=\pm\frac{1}{2}\pi R_7\\(p_8,p_9)\in\Gamma_{2,2}(R_8,R_9)}}\C\ket{x_6,x_7,p_8,p_9}\right).\\
\end{aligned}
\end{equation}

The actions of $g,h$ are given by

\begin{equation}
\begin{aligned}
    g\ket{x_6,x_7,p_8,p_9}=&\ket{-x_6,-x_7,-p_8,-p_9},\\
    h\ket{x_6,x_7,p_8,p_9}=&(-1)^{R_8p_8}\ket{-x_6,-x_7+\pi R_7,p_8,p_9},\\
    gh\ket{x_6,x_7,p_8,p_9}=&(-1)^{R_8p_8}\ket{x_6,x_7-\pi R_7,-p_8,-p_9}.
\end{aligned}
\end{equation}

\begin{equation}
    \begin{aligned}
        \pi R_8=&X^8(2\pi)-X^8(0)\\
        =&\int^{2\pi}_0 d\sigma \partial_\sigma X^8\\
        =&\pi(p_L-p_R)\\
        =&2\pi w_8 R_8.
    \end{aligned}
\end{equation}

Then the winding number of $X^8$ satisfies:

\begin{equation}
    w_8\in \Z+\frac{1}{2},
\end{equation}

which corresponds to the fact $\pi_1(\text{Enriques})=\Z_2$. 

Canonical quantization gives

    \begin{equation}
        \begin{aligned}
            [\alpha^i_r,\alpha^j_s]=&[\tilde{\alpha}^i_r,\tilde{\alpha}^j_s]=r\delta^{ij}\delta_{r+s,0},\\
            [\alpha^i_r,\tilde{\alpha}^j_s]=&0, \quad r,s\in\Z+\frac12,i,j=6,7.
        \end{aligned}
    \end{equation}

 \begin{equation}
        \begin{aligned}
            [\alpha^i_n,\alpha^j_m]=&[\tilde{\alpha}^i_n,\tilde{\alpha}^j_m]=n\delta^{ij}\delta_{n+m,0},\\
            [\alpha^i_n,\tilde{\alpha}^j_m]=&0, \quad n,m\in\Z,i,j=8,9.
        \end{aligned}
    \end{equation}

    $L_0$ is given by

    \begin{equation}
        \begin{aligned}
            L_0=&\frac{1}{2}\sum_{i=6}^7\sum_{r=1/2}\qty(\alpha_{-r}^i\alpha_{r}^i+\alpha_r^i\alpha_{-r}^i)+\frac{1}{2}\sum_{i=8}^9\sum_{n=1}\qty(\alpha_{-n}^i\alpha_{n}^i+\alpha_n^i\alpha_{-n}^i)\\
            =&\sum_{i=6}^7\sum_{r=1/2}\alpha_{-r}^i\alpha_{r}^i+\sum_{i=8}^9\sum_{n=1}\alpha_{-n}^i\alpha_{n}^i-\frac{1}{24}.
        \end{aligned}
    \end{equation}
The results are
\begin{equation}
    \begin{aligned}
        \Gamma_{4,4}^{1,0}\chvec{0}{0}=&\tr_{\H^{h}_{\Gamma_{4,4}}}\q\\
        =&4\Biggl|\frac{1}{q^{\frac{1}{24}}\prod_{n=1}^\infty(1-q^{n-\frac{1}{2}})^2(1-q^n)^2}\Biggr|^2\sum_{n\in\Z}\sum_{w\in \Z+\frac{1}{2}}q^{\frac{1}{2}\qty(\frac{n}{R_8}+wR_8)^2}\bar q^{\frac{1}{2}\qty(\frac{n}{R_8}-wR_8)^2}\vartheta_{\Gamma_{1,1}(R_9)}\\
        =&4\frac{1}{|\eta\vartheta_{01}|^2}\sum_{n\in\Z}\sum_{w\in \Z+\frac{1}{2}}q^{\frac{1}{2}\qty(\frac{n}{R_8}+wR_8)^2}\bar q^{\frac{1}{2}\qty(\frac{n}{R_8}-wR_8)^2}\vartheta_{\Gamma_{1,1}(R_9)},\\
    \Gamma_{4,4}^{1,0}\chvec{0}{1}=&\tr_{\H^{h}_{\Gamma_{4,4}}}g \q=0,\\
        \Gamma_{4,4}^{1,1}\chvec{0}{0}=&\tr_{\H^{h}_{\Gamma_{4,4}}}h \q\\
        =&4\Biggl|\frac{1}{q^{\frac{1}{24}}\prod_{n=1}^\infty(1+q^{n-\frac{1}{2}})^2(1-q^n)^2}\Biggr|^2\sum_{n\in\Z}\sum_{w\in \Z+\frac{1}{2}}(-1)^{R_8 p_8}q^{\frac{1}{2}\qty(\frac{n}{R_8}+wR_8)^2}\bar q^{\frac{1}{2}\qty(\frac{n}{R_8}-wR_8)^2}\vartheta_{\Gamma_{1,1}(R_9)}\\
        =&4\frac{1}{|\eta\vartheta_{00}|^2}\sum_{n\in\Z}\sum_{w\in \Z+\frac{1}{2}}(-1)^nq^{\frac{1}{2}\qty(\frac{n}{R_8}+wR_8)^2}\bar q^{\frac{1}{2}\qty(\frac{n}{R_8}-wR_8)^2}\vartheta_{\Gamma_{1,1}(R_9)},\\
        \Gamma_{4,4}^{1,1}\chvec{0}{1}=&\tr_{\H^{h}_{\Gamma_{4,4}}}gh \q=0.
    \end{aligned}
\end{equation}

\begin{comment}
    \begin{equation}
    \begin{aligned}
        \Gamma_{4,4}^{1,1}\chvec{0}{1}=&\tr_{\H^{h}_{\Gamma_{4,4}}}gh \q
        =\Biggl|\frac{1}{q^{\frac{1}{24}}\prod_{n=1}^\infty(1\blue{-}q^{n-\frac{1}{2}})^2(1+q^n)^2}\Biggr|^2\\
        =&\biggl|\frac{2\eta^2}{\vartheta_{01}\vartheta_{10}}\biggr|^2
    \end{aligned}
\end{equation}
\end{comment}

\subsection{\texorpdfstring{$gh$}{gh}-twisted space}

The boundary conditions of $gh$-twisted bosons $X^i_{gh}:T^2\to T^4$ are

\begin{equation}
    \begin{aligned}
    &X_{gh}^{i}(\sigma_1+2\pi,\sigma_2)\\=&gh\cdot X_{gh}^i(\sigma_1,\sigma_2),~~i=6,7,8,9,
    \end{aligned}
    \end{equation}

    where

     \begin{equation}
             gh :
\begin{cases}
X^6 \rightarrow X^6,\\
X^7 \rightarrow X^7 - \pi R_7 ,\\
X^8 \rightarrow -X^8 - \pi R_8,\\
X^9 \rightarrow -X^9. \\
\end{cases}
 \end{equation}

    After the canonical quantization, we find

\begin{equation}
        \begin{aligned}
            [\alpha^i_n,\alpha^j_m]=&[\tilde{\alpha}^i_n,\tilde{\alpha}^j_m]=n\delta^{ij}\delta_{n+m,0},\\
            [\alpha^i_n,\tilde{\alpha}^j_m]=&0, \quad n,m\in\Z,i,j=6,7,
        \end{aligned}
    \end{equation}

    \begin{equation}
        \begin{aligned}
            [\alpha^i_r,\alpha^j_s]=&[\tilde{\alpha}^i_r,\tilde{\alpha}^j_s]=r\delta^{ij}\delta_{r+s,0},\\
    [\alpha^i_r,\tilde{\alpha}^j_s]=&0, \quad r,s\in\Z+\frac12,i,j=8,9.
        \end{aligned}
    \end{equation}

 and
    \begin{equation}
        \begin{aligned}
            L_0=&\frac{1}{2}\sum_{i=6}^7\sum_{n=1}\qty(\alpha_{-n}^i\alpha_{n}^i+\alpha_n^i\alpha_{-n}^i)+\frac{1}{2}\sum_{i=8}^9\sum_{r=1/2}\qty(\alpha_{-r}^i\alpha_{r}^i+\alpha_r^i\alpha_{-r}^i)\\
            =&\sum_{i=6}^7\sum_{n=1}\alpha_{-n}^i\alpha_{n}^i+\sum_{i=8}^9\sum_{r=1/2}\alpha_{-r}^i\alpha_{r}^i-\frac{1}{24}.
        \end{aligned}
    \end{equation}

Let us denote the Fock space of $X_{gh}^i$ as $\H^{4,4}_{X_{gh}}$.  The $gh$-twisted Hilbert space $\H_{\Gamma_{4,4}}^{gh}$ is given concretely as follows: 

\begin{equation}
\begin{aligned}
\H_{\Gamma_{4,4}}^{gh}=&
\H^{4,4}_{X_{gh}}\otimes \left(\bigoplus_{\substack{ x^8=\pm\frac{1}{2}\pi R_8\\x^9=0,\pi R_9 \\(p_6,p_7)\in\Gamma_{2,2}(R_6,R_7)}}\C\ket{p_6,p_7,x_8,x_9}\right)\\
\end{aligned}
\end{equation}

The actions of $g,h,gh$ are defined as

\begin{equation}
    \begin{aligned}
        g\ket{p_6,p_7,x_8,x_9}=&\ket{-p_6,-p_7,-x_8,-x_9},\\
        h\ket{p_6,p_7,x_8,x_9}=&(-1)^{R_7p_7}\ket{-p_6,-p_7,x_8+\pi R_8,x_9},\\
        gh\ket{p_6,p_7,x_8,x_9}=&(-1)^{R_7p_7}\ket{p_6,p_7,-x_8-\pi R_8,-x_9}.\\
    \end{aligned}
\end{equation}
Then the results are
\begin{equation}
    \begin{aligned}
        \Gamma_{4,4}^{1,0}\chvec{1}{0}=&\tr_{\H^{gh}_{\Gamma_{4,4}}}\q\\
        =&4\Biggl|\frac{1}{q^{\frac{1}{24}}\prod_{n=1}^\infty(1-q^{n-\frac{1}{2}})^2(1-q^n)^2}\Biggr|^2\vartheta_{\Gamma_{1,1}(R_6)}\sum_{n\in\Z}\sum_{w\in \Z+\frac{1}{2}}q^{\frac{1}{2}\qty(\frac{n}{R_7}+wR_7)^2}\bar q^{\frac{1}{2}\qty(\frac{n}{R_7}-wR_7)^2}\\
        =&4\frac{1}{|\eta\vartheta_{01}|^2}\vartheta_{\Gamma_{1,1}(R_6)}\sum_{n\in\Z}\sum_{w\in \Z+\frac{1}{2}}q^{\frac{1}{2}\qty(\frac{n}{R_7}+wR_7)^2}\bar q^{\frac{1}{2}\qty(\frac{n}{R_7}-wR_7)^2},\\
    \Gamma_{4,4}^{1,0}\chvec{1}{1}=&\tr_{\H^{gh}_{\Gamma_{4,4}}}g\q=0,\\
        \Gamma_{4,4}^{1,1}\chvec{1}{0}=&\tr_{\H^{gh}_{\Gamma_{4,4}}}h\q=0,\\
      \Gamma_{4,4}^{1,1}\chvec{1}{1}=&\tr_{\H^{gh}_{\Gamma_{4,4}}}gh\q
         =4\Biggl|\frac{1}{q^{\frac{1}{24}}\prod_{n=1}^\infty(1+q^{n-\frac{1}{2}})^2(1-q^n)^2}\Biggr|^2\\
        =&4\frac{1}{|\eta\vartheta_{00}|^2}\vartheta_{\Gamma_{1,1}(R_6)}\sum_{n\in\Z}\sum_{w\in \Z+\frac{1}{2}}(-1)^nq^{\frac{1}{2}\qty(\frac{n}{R_7}+wR_7)^2}\bar q^{\frac{1}{2}\qty(\frac{n}{R_7}-wR_7)^2}.
    \end{aligned}
\end{equation}
\begin{comment}

\begin{equation}
    \begin{aligned}
        \Gamma_{4,4}^{1,0}\chvec{1}{1}=&\tr_{\H^{gh}_{\Gamma_{4,4}}}g\q\\
         =&\Biggl|\frac{1}{q^{\frac{1}{24}}\prod_{n=1}^\infty(1+q^{n-\frac{1}{2}})^2(1+q^n)^2}\Biggr|^2\\
        =&\biggl|\frac{2\eta^2}{\vartheta_{00}\vartheta_{10}}\biggr|^2
    \end{aligned}
\end{equation}

    \begin{equation}
    \begin{aligned}
        \Gamma_{4,4}^{1,1}\chvec{1}{0}=&\tr_{\H^{gh}_{\Gamma_{4,4}}}h\q\\
         =&\Biggl|\frac{1}{q^{\frac{1}{24}}\prod_{n=1}^\infty(1-q^{n-\frac{1}{2}})^2(1+q^n)^2}\Biggr|^2\\
        =&\biggl|\frac{2\eta^2}{\vartheta_{01}\vartheta_{10}}\biggr|^2
    \end{aligned}
\end{equation}
\end{comment}

\section{6d Supersymmetry}
In this section we review the supermultiplets in six dimensions. In the light-cone formalism, $\Spin(1,5)$ is reduced to $\Spin(0,4)\simeq \SU(2)\times\SU(2)$.

\subsection{\texorpdfstring{$\N=(1,1)$}{N=(1,1)} supermultiplets}

The basic representation of 6d $\N=(1,1)$ supersymmetry is:

\begin{equation}
    \mathcal{R}=\ket{\tfrac{1}{2},\tfrac{1}{2}}+\ket{\tfrac{1}{2},0}^2+\ket{0,\tfrac{1}{2}}^2+\ket{0,0}^4.
\end{equation}
This is called the $6d$ vector multiplet. 

The 6d $\N=(1,1)$ supergravity multiplet is
\begin{equation}
    \begin{aligned}
        \mathcal{R}\otimes\ket{\tfrac{1}{2},\tfrac{1}{2}}=&\ket{1,1}_{G_{\mu\nu}}+\ket{1,0}_{B_{\mu\nu}^+}+\ket{0,1}_{B^-_{\mu\nu}}+\ket{0,0}_{\phi}+\ket{\tfrac{1}{2},\tfrac{1}{2}}^4\\
        +&\ket{1,\tfrac{1}{2}}+\ket{\tfrac{1}{2},1}+\ket{0,\tfrac{1}{2}}+\ket{\tfrac{1}{2},0}.
    \end{aligned}
\end{equation}

\subsection{\texorpdfstring{$\N=(2,0)$}{N=(2,0)} supermultiplets}
The basic representation of 6d $\N=(2,0)$ supersymmetry is:
\begin{equation}
    \mathcal{R}^\prime=\ket{1,0}^2+\ket{\tfrac{1}{2},0}^4+\ket{0,0}^5.
\end{equation}
This is called the 6d $\N=(2,0)$ tensor multiplet. The 6d $\N=(2,0)$ supergravity multiplet is 
\begin{equation}
\mathcal{R}^\prime\otimes \ket{0,1}=\ket{1,1}+\ket{\tfrac{1}{2},1}^4 +\ket{0,1}^5,
\end{equation}
where there are no scalar fields.

\begin{comment}

\section{Mathematical Preliminaries of K3 and Enriques Surfaces}
\subsection{K3 Surfaces}

A K3 surface is a four-dimensional manifold, which preserves one half of supersymmetry. 
\begin{equation}
   \begin{array}{ccccc}
  &   & h^{2,2} &   &   \\
  & h^{2,1} &    & h^{1,2} &   \\
h^{2,0} &   & h^{1,1} &   & h^{0,2} \\
  & h^{1,0} &    & h^{0,1} &   \\
  &   & h^{0,0}  &   &
\end{array}= \begin{array}{ccccc}
  &   & 1  &   &   \\
  & 0 &    & 0 &   \\
1 &   & 20 &   & 1 \\
  & 0 &    & 0 &   \\
  &   & 1  &   &
\end{array}
\end{equation}

\begin{equation}
\begin{aligned}
        H_2(K3;\Z)=&\Gamma_{3,19},\\
        H_0\oplus H_2\oplus H_4=&\Gamma_{4,20}
\end{aligned}
\end{equation}

\subsection{Enriques Surfaces}

\begin{equation}
   \begin{array}{ccccc}
  &   & h^{2,2} &   &   \\
  & h^{2,1} &    & h^{1,2} &   \\
h^{2,0} &   & h^{1,1} &   & h^{0,2} \\
  & h^{1,0} &    &h^{0,1} &   \\
  &   & h^{0,0}  &   &
\end{array}=  \begin{array}{ccccc}
  &   & 1  &   &   \\
  & 0 &    & 0 &   \\
0 &   & 10 &   & 0 \\
  & 0 &    & 0 &   \\
  &   & 1  &   &
\end{array}
\end{equation}

\begin{equation}
    H_2(E;\Z)=\Gamma_{1,9}+\Z/2\Z
\end{equation}

\end{comment}

\end{appendix}

\bibliographystyle{JHEP}
\bibliography{reference}
\end{document}